\documentclass[reprint,superscriptaddress,amsmath,amssymb]{revtex4-2}

\usepackage{amsmath,mathtools,amssymb,bm}
\usepackage{graphicx}
\usepackage[T1]{fontenc}
\usepackage[dvipsnames]{xcolor}
\usepackage[colorlinks=true,linkcolor=blue,citecolor=blue,urlcolor=blue]{hyperref}
\usepackage{ragged2e}
\usepackage{placeins}
\usepackage{orcidlink}

\newcommand{\Tr}{\operatorname{Tr}}
\newcommand{\figpanel}[2]{\hyperref[#1]{\ref*{#1}(#2)}}
\graphicspath{{Figs/}}
\ifdefined\BAclean
  
\else
  
\fi
\begin{document}

\title{Many-Body Flux Selection in Chiral Quantum Batteries}

\author{Borhan Ahmadi\orcidlink{0000-0002-2787-9321}}
\email{borhan.ahmadi@ug.edu.pl}
\affiliation{International Centre for Theory of Quantum Technologies, University of Gda\'nsk, ul. prof. Marii Janion 4, 80-309 Gda\'nsk, Poland}

\begin{abstract}
We study an interacting quantum-battery chain in which a synthetic flux controls interference between hopping paths. Keeping the internal battery Hamiltonian fixed, we seek the flux that leaves the most extractable work. Even a Gaussian reconstruction supplied with exact one-body information can rank competing flux settings incorrectly. A reduced many-body model that retains correlations within a few battery-occupation sectors recovers the correct ranking and selects nearly optimal fluxes across the finite systems and parameter ranges studied. Strikingly, the model can make an accurate control choice even when it excludes configurations carrying substantial probability in the exact state.
Thus, choosing a nearly optimal charging flux can require much less information than reconstructing
the full many-body state.
\end{abstract}

\maketitle

\textit{Introduction.}---
Quantum batteries store energy, but their usefulness also depends on how much of that energy can be extracted as work \cite{Campaioli2024Colloquium,2jtp-jpkn,ferraro2026opportunities,rxf5-zyd8,ahmadi2026Amplifier}. The maximum work extractable by a unitary operation on the battery is its ergotropy; the energy left after extraction is passive \cite{Allahverdyan2004Work}. Coherence and correlations created during charging can change this balance \cite{Andolina2019Extractable,Shi2022Extractable,Cenedese2026Monitoring}. Correlations can also affect how ergotropy is transferred between systems \cite{Simon2025ErgotropyTransport}. A charging protocol must therefore shape the state left in the battery, rather than merely maximize the energy delivered to it.

Interference offers a way to guide this process. Excitations can reach the battery along different paths, and their relative phases determine whether the paths reinforce or oppose one another. Directional transport, chiral couplings, and squeezing have been used to improve charging \cite{Ahmadi2024Nonreciprocal,Ahmadi2026ChiralSqueeze,Yan2026GiantAtom,Liu2026ChiralQB}. Chirality and network topology can affect ergotropy \cite{Cavazzoni2026QuantumWalks}, while topological waveguides can protect energy transfer and stored work \cite{Lu2025Topological}. Studies of many-body batteries also show how interactions and correlations change charging and work extraction \cite{Andolina2019ManyBody,Rossini2019Localized}. Interactions reshape phase-dependent charging \cite{Verma2026Andreev,Zhang2024DM}. Optimal charging controls have been explored in other settings \cite{Mazzoncini2023Optimal}. Yet the phase that loads the most energy need not leave the most extractable work. How do correlations created during charging change the phase that maximizes this work?

This question is especially relevant to lattices with triangular hopping paths. A synthetic flux tunes interference along selected triangles, while the hard-core constraint and intersite interactions make the motion depend on which sites are occupied. Related triangular flux ladders support chiral phases and interaction-dependent transport \cite{Halati2023Triangular,Halati2025HallHardcore}. In a quantum battery, these same processes also shape the many-body state that determines passive energy \cite{Allahverdyan2004Work}. Following the complete state becomes costly as the system grows. A self-consistent Gaussian theory is simpler because it follows one-body occupations and coherences, but it leaves out correlations that cannot be obtained from these quantities. Even exact one-body information does not generally determine the extractable work of a non-Gaussian battery. The challenge is therefore twofold: to understand which correlated processes select the best flux, and to retain enough of them for optimal control without full many-body evolution.

In this Letter, we address these questions in an
isolated, interacting chiral charger--battery chain.
Exact evolution and time-dependent Hartree--Fock
(TDHF) theory \cite{RingSchuck1980} can favor
different charging settings. The disagreement
between exact evolution and TDHF alone does not
tell us whether TDHF predicts the wrong one-body
motion or misses correlations needed to determine
extractable work. To distinguish these effects,
we reconstruct a Gaussian battery state from exact
one-body information and compare its preferred
flux with the exact result. Across the full flux
period, this reconstruction can still rank
competing maxima incorrectly. We then derive an
exact relation showing how the flux changes
extractable work through both the
battery-occupation probabilities and the
many-body states within each occupation sector.
This leads us to a reduced many-body model that
preserves correlations within a significantly
smaller Hilbert space. Remarkably, the model
recovers the correct ranking of competing flux
branches and predicts nearly optimal charging
phases across the studied finite systems, even
when it cannot accurately reconstruct the full
many-body state.
\begin{figure}[t]
\centering
\includegraphics[width=1\linewidth]{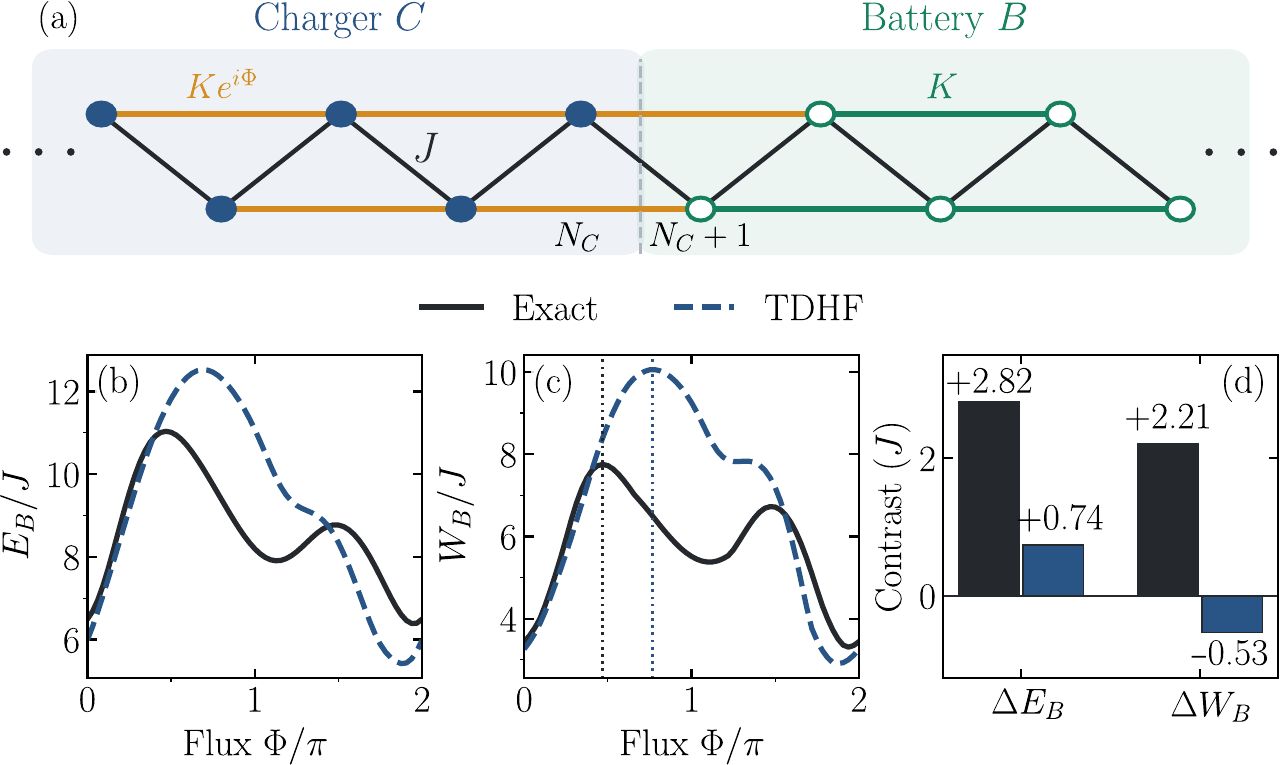}
\caption{\justifying Flux-dependent charging. (a) Filled charger sites and empty battery sites; orange links carry phase $\Phi$, while teal battery links are real. (b) Stored energy and (c) unrestricted ergotropy from exact evolution and TDHF; dotted lines mark the selected fluxes in (c). (d) Energy and ergotropy at $\pi/2$ minus their values at $\pi$. Here $N_C=N_B=10$, $K=0.65J$, $V=2J$, $\omega_C=J$, $\omega_B=4J$, and $Jt=4$.}
\label{fig:flux_failure}
\end{figure}

\textit{Model and observables.}---
We consider an open chain of $N=N_C+N_B$ hard-core-boson sites. The first $N_C$ sites form the charger $C$ and the remaining $N_B$ form the battery $B$. Initially, the charger is filled and the battery is empty [Fig.~\figpanel{fig:flux_failure}{a}]. The operator $b_j$ removes an excitation from site $j$, and $n_j=b_j^\dagger b_j$ counts its occupation, which is either zero or one. The site energy is $\omega_j=\omega_C$ in the charger and $\omega_B$ in the battery. Nearest- and next-nearest-neighbor hopping amplitudes are $J$ and $K$, and neighboring occupations interact with strength $V$. The next-nearest-neighbor phase is $\phi_j=\Phi$ for $j\le N_C$ and zero otherwise. We set $\hbar=1$ and write H.c. for the Hermitian conjugate. The Hamiltonian is
\begin{equation}
\begin{aligned}
H(\Phi)={}&\sum_{j=1}^{N}\omega_j n_j
+V\sum_{j=1}^{N-1}n_jn_{j+1}\\
&+J\sum_{j=1}^{N-1}(b_j^\dagger b_{j+1}+\mathrm{H.c.})\\
&+K\sum_{j=1}^{N-2}(e^{i\phi_j}b_j^\dagger b_{j+2}+\mathrm{H.c.}).
\end{aligned}
\label{eq:full_hamiltonian}
\end{equation}
The hopping links form triangles whose interference is controlled by $\Phi$, defined as the phase accumulated around each selected triangle. The phase-bearing links include those crossing the charger--battery boundary, but all internal battery links are real. The internal Hamiltonian $H_B$ contains only terms entirely within the battery and is therefore independent of $\Phi$ (see Note~\ref{note:sm_hamiltonian} of SM \cite{SuppMat}).

The full state $|\psi(t,\Phi)\rangle$ evolves unitarily. Tracing out the charger gives $\rho_B(t,\Phi)=\Tr_C[|\psi(t,\Phi)\rangle\langle\psi(t,\Phi)|]$. Let $p_a$ be its eigenvalues in descending order and $\varepsilon_a$ the eigenvalues of $H_B$ in ascending order. The stored energy $E_B$, passive energy $E_{\mathrm{pass}}=\sum_a p_a\varepsilon_a$, and standard unrestricted ergotropy $W_B$ satisfy \cite{Allahverdyan2004Work}
\begin{equation}
\begin{aligned}
E_B(t,\Phi)&=\Tr[\rho_B(t,\Phi)H_B],\\
W_B(t,\Phi)&=E_B(t,\Phi)-\sum_a p_a(t,\Phi)\varepsilon_a.
\end{aligned}
\label{eq:observables}
\end{equation}
The sorting includes all battery excitation sectors: unrestricted extraction allows any battery unitary, including one that changes excitation number (see Note~\ref{note:sm_exact} of SM \cite{SuppMat} for further detail). We study this property of the charged battery, not the net work of a complete cycle. Note~\ref{note:sm_operational} of SM \cite{SuppMat} states this scope and compares extraction restricted to number-conserving unitaries.

\textit{Different flux choices from Gaussian dynamics.}---
Figure~\figpanel{fig:flux_failure}{b} compares exact evolution with TDHF for the parameters in the caption. Both use the same initial state and Hamiltonian. Their stored energies differ, and their ergotropies select different fluxes [Fig.~\figpanel{fig:flux_failure}{c}]: $\Phi_*^{\mathrm{ex}}\simeq0.4689\pi$ for exact evolution and $\Phi_*^{\mathrm{HF}}\simeq0.7687\pi$ for TDHF.
The difference also changes which of two fixed settings is preferable. Define $\Delta E_B=E_B(\pi/2)-E_B(\pi)$ and $\Delta W_B=W_B(\pi/2)-W_B(\pi)$. Exact dynamics gives $\Delta E_B=2.818J$ and $\Delta W_B=2.208J$, whereas TDHF gives $0.737J$ and $-0.531J$, respectively, [Fig.~\figpanel{fig:flux_failure}{d}]. Both predict more stored energy at $\pi/2$. However, TDHF predicts a passive-energy increase of $1.268J$, larger than its stored-energy increase. The exact passive-energy increase is only $0.610J$. The two descriptions therefore rank the extractable work at these settings in opposite order. All flux searches cover $[0,2\pi)$; no symmetry under $\Phi\mapsto-\Phi$ is assumed. Grid refinement checks the reported maxima but does not prove a continuous global optimum (see Note~\ref{note:sm_accuracy} of SM \cite{SuppMat}). We then ask why the flux choices differ.

\textit{What one-body information misses.}---
The triangular paths make hopping sensitive to nearby occupations. Under the Jordan--Wigner transformation to fermions $c_j$, a hop over two sites becomes
\cite{JordanWigner1928,LiebSchultzMattis1961}
\begin{equation}
b_j^\dagger b_{j+2}=c_j^\dagger(1-2n_{j+1})c_{j+2}.
\label{eq:jw_string}
\end{equation}
The factor changes sign when the intermediate site is occupied.
Thus $K\ne0$ gives occupation-dependent fermionic motion even at $V=0$. The quadratic limit $K=V=0$ provides a check in which exact and Gaussian evolution agree (see Note~\ref{note:sm_gaussian} of SM \cite{SuppMat}).
Let $C_{ij}=\langle c_j^\dagger c_i\rangle$ be the one-body matrix: its diagonal entries are occupations and its off-diagonal entries are coherences. TDHF expresses higher-order averages through this matrix. Its self-consistent one-body Hamiltonian $F_{\mathrm{HF}}(C)$ includes both $K$ and $V$. The exact equation has an additional term $\mathcal I_{\mathrm{corr}}$ from correlations not fixed by $C$ \cite{RingSchuck1980,Tohyama2020}:
\begin{equation}
\dot C=-i[F_{\mathrm{HF}}(C),C]+\mathcal I_{\mathrm{corr}}.
\label{eq:correlation_source}
\end{equation}
This term vanishes initially but grows during interacting evolution, including at $V=0$ when $K\ne0$ (see Note~\ref{note:sm_source} of SM \cite{SuppMat} for derivation). Its size measures a departure from TDHF, but does not determine the best flux.

To separate errors in $C$ from errors in the state reconstructed from it, let $C_B$ be the battery part of $C$. The number-conserving fermionic Gaussian state $\mathcal G(C_B)$ has this one-body matrix and fixes higher-order averages by Wick's rule. Write its ergotropy as $W_G(C_B)$, evaluated with the same full $H_B$. Using the exact and TDHF matrices, $C_B^{\mathrm{ex}}$ and $C_B^{\mathrm{HF}}$, gives
\begin{equation}
\begin{aligned}
W_B(\rho_B)-W_G(C_B^{\mathrm{HF}})={}&
\big[W_G(C_B^{\mathrm{ex}})-W_G(C_B^{\mathrm{HF}})\big]\\
&+\big[W_B(\rho_B)-W_G(C_B^{\mathrm{ex}})\big].
\end{aligned}
\label{eq:gaussian_ledger}
\end{equation}
The first bracket changes only the one-body input to the Gaussian calculation. The second compares the exact and Gaussian states with the same input. These are parts of one difference, not separate sources of work. We note that the exact-data reconstruction requires full many-body information and is a diagnostic, not an independent predictor (see Note~\ref{note:sm_reconstruction} of SM \cite{SuppMat} for more details).

We compare it with a reduced model that keeps many-body correlations. Let $q$ be the instantaneous number of battery excitations and $P_{\le m}$ project onto configurations with $q\le m$. We note that a sector boundary is not a boundary between two physical sites, it is a boundary between two groups of many-body configurations. This limits battery occupation, not the number of hopping events. The resulting configuration-interaction (CI) model evolves the initial state as
\begin{equation}
\begin{aligned}
H_m(\Phi)&=P_{\le m}H(\Phi)P_{\le m},\\
|\psi_m(t,\Phi)\rangle&=e^{-itH_m(\Phi)}|\psi(0)\rangle.
\end{aligned}
\label{eq:ci}
\end{equation}
The projected Hamiltonian is Hermitian and retains correlations within the allowed space. It omits paths through higher occupations. Its reduced battery state is evaluated with the unchanged full $H_B$, without restricting extraction to that space. Unlike the exact-data Gaussian construction, it selects a phase without a full exact trajectory (see Note~\ref{note:sm_ci} of SM \cite{SuppMat} for further details).

For the initially filled charger, the retained dimension is $D_m=\sum_{q=0}^m\binom{N_C}{q}\binom{N_B}{q}$. At fixed $m$ in balanced chains, it grows as $O(N^{2m})$, rather than exponentially as in exact evolution. This reduces the state-propagation space, but does not guarantee a size-independent cutoff or efficient evaluation of unrestricted work (see Note~\ref{note:sm_ci} of SM \cite{SuppMat}).

\begin{figure}[t]
\centering
\includegraphics[width=1\linewidth]{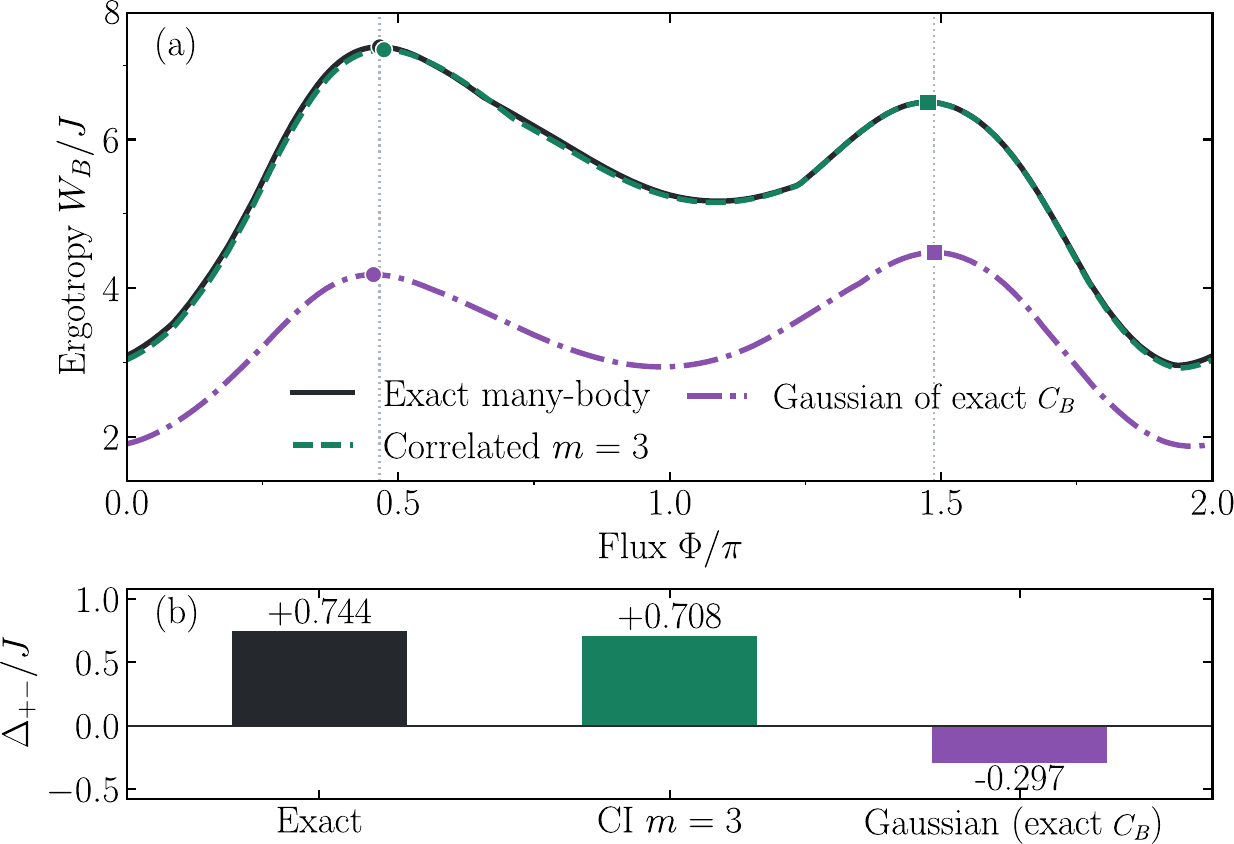}
\caption{\justifying Exact one-body data can still give the wrong flux choice. (a) Unrestricted ergotropy from exact evolution, Gaussian reconstruction of exact $C_B$, and correlated CI with $m=3$. Circles and squares mark positive- and negative-branch maxima; dotted lines mark the exact and Gaussian selections. (b) Difference between the two branch maxima for each method. Gaussian reconstruction favors the wrong branch; CI recovers the exact ranking. Here $N_C=N_B=8$; other parameters are those of Fig.~\ref{fig:flux_failure}.}
\label{fig:fullperiod}
\end{figure}

Figure~\figpanel{fig:fullperiod}{a} compares the three ergotropy curves at $N_C=N_B=8$, $V=2J$, and $Jt=4$. Exact evolution selects $0.4658\pi$. Gaussian reconstruction from exact $C_B$ has a nearby local maximum at $0.4545\pi$, but a higher maximum at $1.4876\pi$, equivalent to $-0.5124\pi$. It therefore selects the opposite half of the flux period despite its exact one-body input. The independently evolved $m=3$ model selects $0.473872\pi$, close to the exact choice.

Let us call $[0,\pi)$ the positive branch and $[\pi,2\pi)$ the negative branch. For each method, let $W_+^*$ and $W_-^*$ be its largest ergotropies on these branches. Figure~\figpanel{fig:fullperiod}{b} compares $\Delta_{+-}=W_+^*-W_-^*$: exact evolution gives $+0.7441J$, Gaussian reconstruction $-0.2971J$, and $m=3$ gives $+0.7077J$. Thus the reduced model recovers both the sign and approximate size of the branch preference. To assess a chosen phase, we use its \emph{exact ergotropy regret}, defined as the difference between the maximum exact ergotropy and the exact ergotropy obtained at that flux. The Gaussian and $m=3$ choices lose $0.7483J$ and $0.0018J$, respectively.
The ranking depends on both stored and passive energy. At the same two Gaussian-selected peaks, their exact positive-minus-negative differences are $2.1977J$ and $1.4537J$. Gaussian reconstruction instead gives $1.8531J$ and $2.1502J$: its passive-energy increase exceeds its stored-energy increase and reverses the work preference. This reversal occurs despite using exact one-body information, pointing to the role of many-body correlations in determining the balance between stored and passive energy.

\textit{How higher occupations affect the flux choice.}---
Number conservation makes $\rho_B$ block diagonal in $q$: $\rho_B=\bigoplus_qP_q\sigma_q$, where $P_q$ is the occupation probability and $\sigma_q$ the normalized conditional state. Let $H_B^{(q)}$ be the battery Hamiltonian in that sector. To differentiate passive energy, define $\mathcal K_q$ in the eigenbasis of $\sigma_q$: it assigns each eigenvector an energy of the full $H_B$ according to the global rank of its eigenvalue in $\rho_B$, not its rank within $q$. Wherever passive energy is differentiable,
\begin{equation}
\begin{aligned}
\partial_\Phi W_B={}&\sum_q(\partial_\Phi P_q)
\Tr[(H_B^{(q)}-\mathcal K_q)\sigma_q]\\
&+\sum_qP_q\Tr[(H_B^{(q)}-\mathcal K_q)
\partial_\Phi\sigma_q].
\end{aligned}
\label{eq:sector_response}
\end{equation}
The first term describes changes in occupation probabilities; the second describes changes in the states within those sectors. Both can matter, so occupation probabilities alone need not locate the optimum. This is an exact response identity, not a closed prediction rule (see Note~\ref{note:sm_reconstruction} of SM \cite{SuppMat} for more details).
Each hop across the charger--battery boundary changes $q$ by one. To examine the effect of access to higher occupations, let $B_q(\Phi)$ be the Hermitian coupling between sectors $q$ and $q+1$. We change only this coupling, leaving all other terms and the full $H_B$ fixed:
\begin{equation}
H_q(\lambda,\Phi)=H(\Phi)-(1-\lambda)B_q(\Phi),
\quad 0\le\lambda\le1.
\label{eq:gate}
\end{equation}
At $\lambda=0$, the initially empty battery cannot reach occupations above $q$; $\lambda=1$ restores the original Hamiltonian. This is a diagnostic modification, not an assumed experimental control (see Note~\ref{note:sm_transfer} of SM \cite{SuppMat}).

\begin{figure}[t]
\centering
\includegraphics[width=1\linewidth]{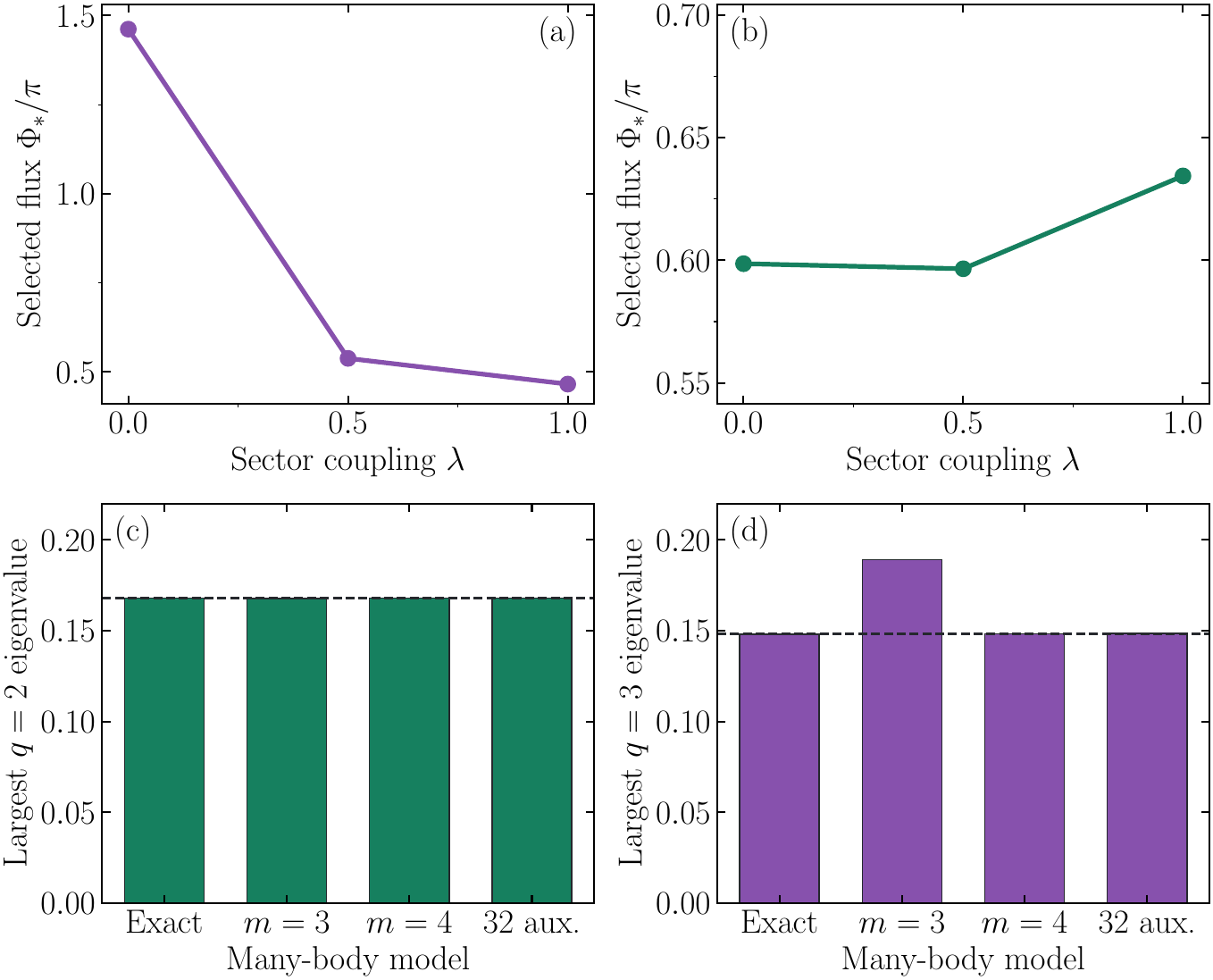}
\caption{\justifying Effect of higher battery occupations. Selected flux as (a) the $q=1\leftrightarrow2$ coupling is restored at $V=2J$ and (b) the $q=2\leftrightarrow3$ coupling is restored at $V=1.5J$; $N=16$, $Jt=4$. Lines join the three tested strengths. Largest reduced-state eigenvalue in (c) $q=2$ and (d) $q=3$, comparing exact evolution, cutoffs $m=3,4$, and $m=3$ with 32 auxiliary vectors from $q=4$. Dashed lines give exact values; $N=20$, $V=2J$, $Jt=4$, $\Phi=0.4689\pi$. Other parameters are those of Fig.~\ref{fig:flux_failure}.}
\label{fig:feedback}
\end{figure}

For $N=16$, $V=2J$, and $Jt=4$, restoring the $q=1\leftrightarrow2$ coupling changes the selected phase from $1.46183\pi$ at zero strength to $0.53797\pi$ at half strength and $0.46577\pi$ at full strength [Fig.~\figpanel{fig:feedback}{a}]. Access to $q=2$ therefore changes the preferred branch. At $V=1.5J$, restoring the $q=2\leftrightarrow3$ coupling gives $0.59860\pi$, $0.59653\pi$, and $0.63438\pi$ [Fig.~\figpanel{fig:feedback}{b}]. The change is not monotonic: the coupling permits motion in both directions and also changes amplitudes returning to lower occupations.
Higher occupations also affect the passive-energy ordering. Figures~\figpanel{fig:feedback}{c} and \figpanel{fig:feedback}{d} compare the largest reduced-state eigenvalues in sectors $q=2$ and $q=3$ at $N=20$. The $m=3$ model nearly reproduces the $q=2$ value, but raises the $q=3$ value from $0.14811$ to $0.18919$, above the leading $q=2$ eigenvalue. Adding the $q=4$ sector, or 32 auxiliary vectors from it, restores their exact order. Because passive energy uses the globally sorted eigenvalues, this error affects work even in retained sectors. It does not establish that this rank change alone causes the flux shift (see Note~\ref{note:sm_transfer} of SM \cite{SuppMat} for further details).

\begin{figure}[t]
\centering
\includegraphics[width=1\linewidth]{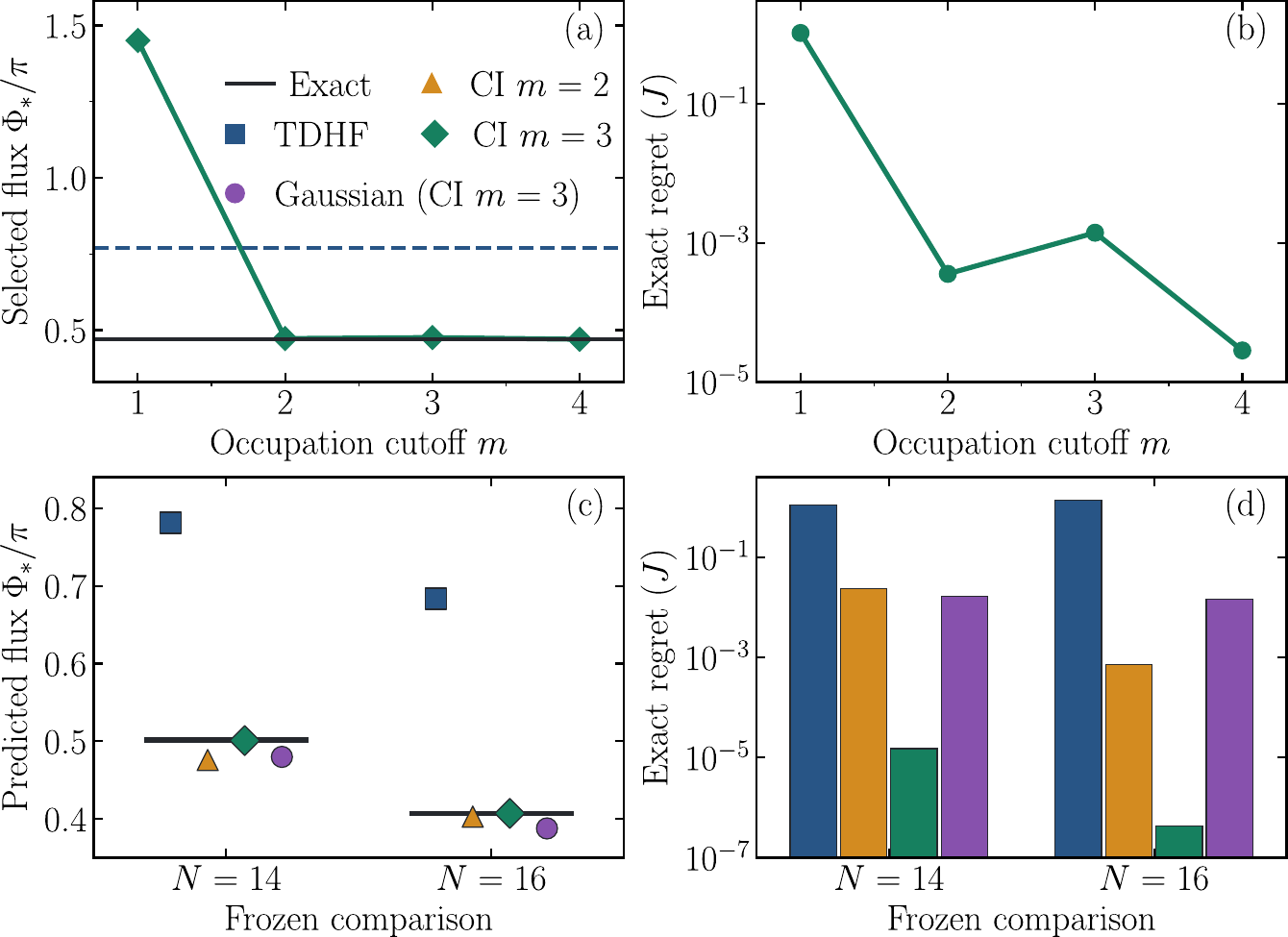}
\caption{\justifying Control accuracy of the reduced model. (a) Selected full-period flux and (b) exact ergotropy regret versus cutoff $m$ at $N=20$, $V=2J$, $Jt=4$. (c) Predictions saved before exact evaluation and (d) their exact regrets for $N=14$, $V=1.8J$, $Jt=4.6$ and $N=16$, $V=2.3J$, $Jt=4.35$. The green curve in (a),(b) varies $m$; green symbols and bars in (c),(d) use $m=3$. Panels (c),(d) retain the original positive-branch choices; horizontal marks in (c) give exact references. Gaussian (CI $m=3$) reconstructs a Gaussian state from the projected one-body matrix. Other parameters are those of Fig.~\ref{fig:flux_failure}.}
\label{fig:prediction}
\end{figure}

\textit{Predicting the optimal charging flux.}---
We now examine how much of this many-body structure is needed for optimal control of charging. Figure~\figpanel{fig:prediction}{a} varies the cutoff in Eq.~\eqref{eq:ci} at $N=20$. The $m=1$ model selects $1.4507\pi$. Allowing $q\le2$ moves the choice to $0.4723\pi$, close to the exact $0.4689\pi$, while retaining only $2126$ of $184756$ configurations. The $m=3$ and $m=4$ choices are $0.4755\pi$ and $0.4698\pi$. The exact regret is $3.58\times10^{-4}J$ at $m=2$ and $1.40\times10^{-3}J$ at $m=3$, before decreasing at $m=4$ [Fig.~\figpanel{fig:prediction}{b}].
Although the retained spaces are nested, there is no variational principle ordering the selected flux or its regret as $m$ increases. For a smooth, nondegenerate exact maximum with a nearby projected maximum, the leading local regret is
\begin{equation}
\mathcal R_m\simeq
\frac{[\partial_\Phi\delta_m(\Phi_*^{\mathrm{ex}})]^2}
{2|\partial_\Phi^2 W_B^{\mathrm{ex}}(\Phi_*^{\mathrm{ex}})|},
\label{eq:local_regret}
\end{equation}
where $\delta_m(\Phi)=W_B^{(m)}(\Phi)-W_B^{\mathrm{ex}}(\Phi)$ is the work error (see Note~\ref{note:sm_ci} of SM~\cite{SuppMat} for derivation). Thus local regret depends on the flux variation of this error, not simply its magnitude. The smaller $m=2$ regret at this point is not systematic: it selects the wrong branch at five points of the interaction--time scan below, whereas $m=3$ selects the correct branch throughout. The local relation does not describe such distant branch switches.

This control accuracy does not imply state accuracy. At the exact reference phase, the probability outside $q\le2$ is $P_{>2}\simeq0.416$. The squared overlap of the exact global state with any normalized state in that subspace is therefore at most $1-P_{>2}\simeq0.584$. A small control regret can coexist with this limited state overlap because it measures the location of a maximum, not agreement between states at the same flux.

The separation also concerns the reduced battery state itself. Writing $\rho_B^{(m)}=\Tr_C[|\psi_m\rangle\langle\psi_m|]$ and $P_{>m}$ for the exact probability of battery occupations above $m$, the trace-distance variational principle gives (see
Note~\ref{note:sm_ci} of SM~\cite{SuppMat} for derivation)
\begin{equation}
\frac12\|\rho_B-\rho_B^{(m)}\|_1\ge P_{>m}.
\label{eq:reduced_battery_trace_distance}
\end{equation}
Indeed, the projected state has no weight in the excluded sectors, which can be distinguished by measuring battery occupation. At the reference phase this gives a trace distance of at least $0.416$ for $m=2$, despite its small exact control regret.
The occupation hierarchy also gives a rigorous control-error bound. Because each interface hop changes battery occupation by one, the projected state can leak into the excluded space only through the $q=m\leftrightarrow m+1$ boundary. Writing $\epsilon_m=\sup_\Phi|W_B^{(m)}-W_B^{\mathrm{ex}}|$, we obtain
\begin{equation}
\begin{aligned}
\mathcal R_m
&\le 2\epsilon_m
\le 4\Delta_B\sup_\Phi\beta_m(t,\Phi),\\
\beta_m(t,\Phi)
&=\int_0^t ds\,
\|B_m(\Phi)\psi_m(s,\Phi)\|_2.
\end{aligned}
\label{eq:cutoff_convergence}
\end{equation}
Here $\Delta_B$ is the spectral width of the full battery Hamiltonian, $B_m$ couples the last retained sector to the first excluded one, and $\psi_m$ is the independently evolved projected state. This bound requires only the projected dynamics,
although it can be conservative (see
Note~\ref{note:sm_ci} of SM~\cite{SuppMat} for derivation).

Remarkably, numerical convergence of the work curve need not translate into monotonic control accuracy. At $N=16$, increasing $m$ from $3$ to $4$ reduces the largest sampled work-curve error by about one order of magnitude, while the exact regret falls by more than five orders
of magnitude, reaching $2.96\times10^{-9}J$. To investigate size dependence at fixed $K=0.65J$, $V=2J$, and $Jt=4$, we also compare balanced chains with $N=12,14,16,18,20$. For $m=3$, the retained fraction falls from $662/924$ to $16526/184756$, yet the selected flux stays on the exact branch and its regret remains below $1.9\times10^{-3}J$ at all five sizes. This controlled finite-size result does not establish a universal cutoff (see Note~\ref{note:sm_ci} of SM \cite{SuppMat} for further details).

Next we examine choices saved before the exact references were computed. Figures~\figpanel{fig:prediction}{c} and \figpanel{fig:prediction}{d} show two such cases. The $m=3$ selections are $0.5009\pi$ and $0.4072\pi$, close to the exact $0.5015\pi$ and $0.4073\pi$. Their exact regrets are $1.52\times10^{-5}J$ and $4.33\times10^{-7}J$, compared with $1.113J$ and $1.375J$ for TDHF. Figures~\figpanel{fig:prediction}{c} and \figpanel{fig:prediction}{d} retain the original positive-branch comparisons. Subsequent full-period searches confirm the exact and correlated CI choices, but reveal a higher Gaussian-reconstruction maximum at $1.5386\pi$ for $N=16$, with exact regret $0.5362J$. The original Gaussian choice is therefore a local positive-branch optimum, not its full-period optimum.

To examine nearby conditions, we scan $1.5\le V/J\le2.5$ and $3.5\le Jt\le5$ on a $9\times7$ grid at $N_C=N_B=8$. The exact-data Gaussian selects the wrong branch at 35 of 63 points; $m=3$ makes no branch errors, with largest exact regret $0.0034J$ versus $0.8227J$ for the Gaussian. Exact dynamics favors the positive branch throughout this window, so branch agreement alone is insufficient. The \(m=3\) model also selects a flux that gives more exact extractable work than the fixed choice \(\Phi=\pi/2\) in 62 of the 63 cases. The $m=2$ cutoff fails the branch test at five points. The regret and branch maps, including this cutoff comparison, appear in Note~\ref{note:sm_reconstruction} of SM \cite{SuppMat}.

We also vary the phase-bearing hopping $K/J$ from $0.1$ to $1.0$
(eleven sampled values) at fixed $V=2J$ and $Jt=4$.
The exact-data Gaussian selects the wrong branch at every point;
CI with $m=3$ or $m=4$ retains the exact branch throughout, with
largest exact regrets $1.86\times10^{-3}J$ and $1.68\times10^{-4}J$, respectively. The $m=2$ model fails the branch test at two values. Thus the correlated model remains predictive when the hopping that creates the interference is varied (see Note~\ref{note:sm_reconstruction}
of SM~\cite{SuppMat}).

The predictive advantage extends beyond the reference
parameters. Across twelve additional combinations
of hopping strengths, interactions, charging times,
and charger--battery sizes, the correlated $m=3$
model yields a median relative exact regret of
$0.0021\%$, compared with $7.56\%$ for TDHF
(see Note~\ref{note:sm_validation} of
SM~\cite{SuppMat}). Together with the interaction--time
and hopping-strength scans, these results show
that a small correlated occupation space can retain
the information needed for accurate flux selection
across distinct operating conditions.

\textit{Summary and discussion.}---
Selecting the best charging flux requires comparing extractable work, not stored energy alone. We show that even a Gaussian reconstruction using exact one-body information can favor the wrong flux branch. The missing many-body correlations affect the stored and passive energies and can reverse the preferred setting. Our exact occupation-resolved response identity shows how both the probabilities of different battery occupations and the states within these sectors enter the work response. Changing the couplings between sectors further reveals how higher occupations influence the preferred flux.
A reduced model retaining many-body correlations recovers the correct branch ranking and selects nearly optimal fluxes in the finite systems studied. Strikingly, this remains possible even when the model excludes configurations carrying a substantial part of the exact state's probability. The reason is that choosing a good control requires locating the maximum of extractable work, not reconstructing the entire many-body state. The occupation cutoff needed for this accuracy is not universal, however, and its behavior in larger systems remains to be established.

\emph{Acknowledgments}---B.A. acknowledges support from IRA Programme (project no. FENG.02.01-IP.05-0006/23) financed by the FENG program 2021-2027, Priority FENG.02, Measure FENG.02.01., with the support of the FNP.

\textit{Data and code availability}.---
The numerical code, figure data, and records of all twelve additional tests are publicly available in our GitHub repository \cite{ManyBodyFlux}. Further numerical details are given in Note~\ref{note:sm_validation} of SM~\cite{SuppMat}.

\bibliography{References}

@misc{SuppMat,
  author    = {},
  title     = {},
  year      = {},
  note   = {See Supplemental Material following this Letter for the Hamiltonian convention, Gaussian reconstruction, occupation-sector interventions, prediction comparisons, full-period searches, interaction--time robustness, extraction constraints, and numerical stress tests.}
}

@article{ahmadi2026Amplifier,
  title = {{Finite quantum pumps for phase-locked parametric amplification: requirements for quantum-battery work sources}},
  author = {Ahmadi, Borhan},
  journal = {New J. Phys.},
  volume  = {28},
  year    = {2026},
  pages = {104514},
  year = {2026},
  url = {https://doi.org/10.1088/1367-2630/aeb4e6}
}

@article{rxf5-zyd8,
  title = {Quantum-battery-powered geometric Landau-Zener interferometry},
  author = {Ahmadi, Borhan},
  journal = {Phys. Rev. E},
  pages = {},
  year = {2026},
  month = {Nov},
  publisher = {American Physical Society},
  doi = {10.1103/rxf5-zyd8},
  url = {https://link.aps.org/doi/10.1103/rxf5-zyd8}
}

@article{Campaioli2024Colloquium,
  author = {Francesco Campaioli and Stefano Gherardini and James Q. Quach and Marco Polini and Gian Marcello Andolina},
  title = {Colloquium: Quantum batteries},
  year = {2024},
  journal = {Rev. Mod. Phys.},
  volume = {96},
  pages = {031001},
  doi = {10.1103/RevModPhys.96.031001},
  url = {https://doi.org/10.1103/RevModPhys.96.031001}
}

@article{ferraro2026opportunities,
  title={Opportunities and challenges of quantum batteries},
  author={Ferraro, Dario and Cavaliere, Fabio and Genoni, Marco G and Benenti, Giuliano and Sassetti, Maura},
  journal={Nat. Rev. Phys.},
  pages={115–127},
  volume={8},
  year={2026},
  publisher={Nature Publishing Group},
  url={https://doi.org/10.1038/s42254-025-00906-5}
}

@article{2jtp-jpkn,
  title = {Charge-preserving operations in quantum batteries},
  author = {Malavazi, André H. A. and Ahmadi, Borhan and Horodecki, Paweł and Dieguez, Pedro R.},
  journal = {PRX Energy},
  volume = {5},
  pages = {023004},
  year = {2026},
  month = {Mar},
  publisher = {American Physical Society},
  doi = {10.1103/2jtp-jpkn},
  url = {https://doi.org/10.1103/2jtp-jpkn}
}

@article{Allahverdyan2004Work,
  author = {A. E. Allahverdyan and R. Balian and Th. M. Nieuwenhuizen},
  title = {Maximal work extraction from finite quantum systems},
  year = {2004},
  journal = {Europhys. Lett.},
  volume = {67},
  pages = {565-571},
  doi = {10.1209/epl/i2004-10101-2},
  url = {https://doi.org/10.1209/epl/i2004-10101-2}
}

@article{Andolina2019Extractable,
  author = {Gian Marcello Andolina and Maximilian Keck and Andrea Mari and Michele Campisi and Vittorio Giovannetti and Marco Polini},
  title = {Extractable Work, the Role of Correlations, and Asymptotic Freedom in Quantum Batteries},
  journal = {Phys. Rev. Lett.},
  volume = {122},
  pages = {047702},
  year = {2019},
  doi = {10.1103/PhysRevLett.122.047702},
  url = {https://doi.org/10.1103/PhysRevLett.122.047702}
}

@article{Simon2025ErgotropyTransport,
  author = {Rick P. A. Simon and Janet Anders and Karen V. Hovhannisyan},
  title = {Correlations Enable Lossless Ergotropy Transport},
  year = {2025},
  journal = {Phys. Rev. Lett.},
  volume = {134},
  pages = {010408},
  doi = {10.1103/PhysRevLett.134.010408},
  url = {https://doi.org/10.1103/PhysRevLett.134.010408}
}

@article{Ahmadi2024Nonreciprocal,
  author = {Borhan Ahmadi and Pawel Mazurek and Pawel Horodecki and Shabir Barzanjeh},
  title = {Nonreciprocal Quantum Batteries},
  year = {2024},
  journal = {Phys. Rev. Lett.},
  volume = {132},
  pages = {210402},
  doi = {10.1103/PhysRevLett.132.210402},
  url = {https://doi.org/10.1103/PhysRevLett.132.210402}
}

@article{Ahmadi2026ChiralSqueeze,
  author = {Borhan Ahmadi and Andre H. A. Malavazi and Janine Splettstoesser and Pawel Horodecki and Lei Du},
  title = {Charging Quantum Batteries with Chiral Squeezing},
  year = {2026},
  journal = {Phys. Rev. Lett.},
  volume = {137},
  pages = {180404},
  doi = {10.1103/zg87-t4y5},
  url = {https://doi.org/10.1103/zg87-t4y5}
}

@article{Yan2026GiantAtom,
  author = {Ke-Xiong Yan and Yang Liu and Yang Xiao and Jun-Hao Lin and Jie Song and Ye-Hong Chen and Franco Nori and Yan Xia},
  title = {Giant-Atom Quantum Batteries: Lossless Energy Transfer via Interference Engineering},
  year = {2026},
  journal = {Phys. Rev. Lett.},
  volume = {136},
  pages = {180401},
  doi = {10.1103/43n6-rnj3},
  url = {https://doi.org/10.1103/43n6-rnj3}
}

@article{Liu2026ChiralQB,
  author = {Rong-Fang Liu and Wan-Lu Song and Wan-Li Yang and Hua Guan and Jun-Hong An},
  title = {Chiral quantum batteries},
  journal = {arXiv:2603.25173 (2026)},
  url = {https://arxiv.org/abs/2603.25173}
}

@article{Cavazzoni2026QuantumWalks,
  author = {Simone Cavazzoni and Giovanni Ragazzi and Paolo Bordone and Matteo G. A. Paris},
  title = {Quantum Walks as a Tool to Design Robust Quantum Batteries: The Role of Topology and Chirality},
  year = {2026},
  journal = {Adv. Quantum Technol.},
  volume = {9},
  pages = {e00897},
  doi = {10.1002/qute.202500897},
  url = {https://doi.org/10.1002/qute.202500897}
}

@article{Verma2026Andreev,
  author = {Disha Verma and B. Vigneshwar and R. Sankaranarayanan},
  title = {Interaction-Enhanced Ergotropy in Phase-Driven Andreev Bound State Quantum Batteries},
  journal = {arXiv:2606.24456 (2026)},
  url = {https://arxiv.org/abs/2606.24456}
}

@article{Zhang2024DM,
  author = {Xiang-Long Zhang and Xue-Ke Song and Dong Wang},
  title = {Quantum Battery in the Heisenberg Spin Chain Models with Dzyaloshinskii-Moriya Interaction},
  year = {2024},
  journal = {Adv. Quantum Technol.},
  volume = {7},
  pages = {2400114},
  doi = {10.1002/qute.202400114},
  url = {https://doi.org/10.1002/qute.202400114}
}

@article{Mazzoncini2023Optimal,
  author = {Francesco Mazzoncini and Vasco Cavina and Gian Marcello Andolina and Paolo Andrea Erdman and Vittorio Giovannetti},
  title = {Optimal Control Methods for Quantum Batteries},
  journal = {Phys. Rev. A},
  volume = {107},
  pages = {032218},
  year = {2023},
  doi = {10.1103/PhysRevA.107.032218},
  url = {https://doi.org/10.1103/PhysRevA.107.032218}
}

@article{Halati2023Triangular,
  author = {Catalin-Mihai Halati and Thierry Giamarchi},
  title = {Bose-Hubbard triangular ladder in an artificial gauge field},
  year = {2023},
  journal = {Phys. Rev. Research},
  volume = {5},
  pages = {013126},
  doi = {10.1103/PhysRevResearch.5.013126},
  url = {https://doi.org/10.1103/PhysRevResearch.5.013126}
}

@article{Halati2025HallHardcore,
  author = {Catalin-Mihai Halati and Thierry Giamarchi},
  title = {Exploring frustration effects of strongly interacting bosons via the Hall response},
  year = {2025},
  journal = {Phys. Rev. Research},
  volume = {7},
  pages = {013199},
  doi = {10.1103/PhysRevResearch.7.013199},
  url = {https://doi.org/10.1103/PhysRevResearch.7.013199}
  }

@article{Shi2022Extractable,
  author  = {Shi, Hai-Long and Ding, Shu and Wan, Qing-Kun and Wang, Xiao-Hui and Yang, Wen-Li},
  title   = {Entanglement, Coherence, and Extractable Work in Quantum Batteries},
  journal = {Phys. Rev. Lett.},
  volume  = {129},
  pages   = {130602},
  year    = {2022},
  doi     = {10.1103/PhysRevLett.129.130602}
}

@article{Cenedese2026Monitoring,
  author  = {Cenedese, Gabriele and Benenti, Giuliano and Ferraro, Dario and Genoni, Marco G.},
  title   = {Boosting Work Extraction in Quantum Batteries via Continuous Environment Monitoring},
  journal = {Phys. Rev. Lett.},
  volume  = {137},
  pages   = {050406},
  year    = {2026},
  doi     = {10.1103/f591-hx6x}
}

@book{RingSchuck1980,
  author    = {Ring, Peter and Schuck, Peter},
  title     = {The Nuclear Many-Body Problem},
  publisher = {Springer},
  address   = {Berlin},
  year      = {1980},
  url={https://api.semanticscholar.org/CorpusID:120434868}
}

@article{Lu2025Topological,
  author  = {Lu, Zhi-Guang and Tian, Guoqing and L{\"u}, Xin-You and Shang, Cheng},
  title   = {Topological Quantum Batteries},
  journal = {Phys. Rev. Lett.},
  volume  = {134},
  pages   = {180401},
  year    = {2025},
  doi     = {10.1103/PhysRevLett.134.180401}
}

@article{Andolina2019ManyBody,
  author  = {Andolina, Gian Marcello and Keck, Maximilian and Mari, Andrea and Giovannetti, Vittorio and Polini, Marco},
  title   = {Quantum versus classical many-body batteries},
  journal = {Phys. Rev. B},
  volume  = {99},
  pages   = {205437},
  year    = {2019},
  doi     = {10.1103/PhysRevB.99.205437}
}

@article{Rossini2019Localized,
  author  = {Rossini, Davide and Andolina, Gian Marcello and Polini, Marco},
  title   = {Many-body localized quantum batteries},
  journal = {Phys. Rev. B},
  volume  = {100},
  pages   = {115142},
  year    = {2019},
  doi     = {10.1103/PhysRevB.100.115142}
}

@article{JordanWigner1928,
  author  = {P. Jordan and E. Wigner},
  title   = {{\"U}ber das Paulische {\"A}quivalenzverbot},
  journal = {Z. Phys.},
  volume  = {47},
  pages   = {631--651},
  year    = {1928},
  doi     = {10.1007/BF01331938}
}

@article{LiebSchultzMattis1961,
  author  = {E. Lieb and T. Schultz and D. Mattis},
  title   = {Two Soluble Models of an Antiferromagnetic Chain},
  journal = {Ann. Phys. (N.Y.)},
  volume  = {16},
  pages   = {407--466},
  year    = {1961},
  doi     = {10.1016/0003-4916(61)90115-4}
}

@article{Tohyama2020,
  author  = {M. Tohyama},
  title   = {Applications of Time-Dependent Density-Matrix Approach},
  journal = {Front. Phys.},
  volume  = {8},
  pages   = {67},
  year    = {2020},
  doi     = {10.3389/fphy.2020.00067}
}

@article{ManyBodyFlux,
  journal = {https://github.com/Borhan19/ManyBodyFlux},
  url = {https://github.com/Borhan19/ManyBodyFlux?utm_source=chatgpt.com}
  }

\clearpage
\onecolumngrid
\setcounter{section}{0}
\setcounter{subsection}{0}
\setcounter{equation}{0}
\setcounter{figure}{0}
\setcounter{table}{0}
\renewcommand{\thesection}{\arabic{section}}
\renewcommand{\thesubsection}{\Alph{subsection}}
\renewcommand{\theequation}{S\arabic{equation}}
\renewcommand{\thefigure}{S\arabic{figure}}
\renewcommand{\thetable}{S\arabic{table}}
\renewcommand{\theHequation}{SM.\arabic{equation}}
\renewcommand{\theHfigure}{SM.\arabic{figure}}
\renewcommand{\theHtable}{SM.\arabic{table}}
\renewcommand{\theHsection}{SM.\arabic{section}}

\newcommand{\suppnote}[1]{%
\refstepcounter{section}%
\setcounter{subsection}{0}%
\section*{Supplementary Note \arabic{section}. #1}%
}
\newcommand{\suppsubsection}[1]{%
\refstepcounter{subsection}%
\subsection*{\Alph{subsection}. #1}%
}

\section*{Supplemental Material}

We give the model and numerical details in the order needed to reproduce the comparisons. Notes~\ref{note:sm_hamiltonian} and \ref{note:sm_exact} define the Hamiltonians, reduced states, and work calculation. Notes~\ref{note:sm_gaussian} and \ref{note:sm_source} describe the Gaussian approximation and the correlations it omits. Note~\ref{note:sm_reconstruction} compares the resulting flux choices and derives the work-response identities. Notes~\ref{note:sm_transfer} and \ref{note:sm_ci} describe changes to occupation-sector couplings and the reduced many-body model. The last three notes give numerical checks, number-conserving extraction tests, and the complete set of twelve additional parameter tests. Unless marked by the superscript $\mathrm{num}$, ergotropy allows arbitrary battery unitaries and is evaluated with the same full internal battery Hamiltonian.

\suppnote{Hamiltonian, flux convention, and fixed battery Hamiltonian}
\label{note:sm_hamiltonian}

The open chain has $N=N_C+N_B$ sites. Sites $1,\ldots,N_C$ form the charger and sites $N_C+1,\ldots,N$ form the battery. A hard-core-boson operator $b_j$ removes an excitation at site $j$, and $n_j=b_j^\dagger b_j$ has eigenvalues zero and one. Operators on different sites commute, while $(b_j^\dagger)^2=0$. We set $\hbar=1$. The site energy $\omega_j$ is $\omega_C$ in the charger and $\omega_B$ in the battery. In Eq.~\eqref{eq:full_hamiltonian}, $J$ and $K$ are nearest- and next-nearest-neighbor hopping amplitudes, $V$ is the nearest-neighbor interaction, and H.c. denotes the Hermitian conjugate.

The flux $\Phi$ enters the next-nearest-neighbor hopping through the link phase $\phi_j$. For the link indices $1\le j\le N-2$, we use
\begin{equation}
\phi_j=\begin{cases}
\Phi,&1\le j\le N_C,\\
0,&N_C<j\le N-2.
\end{cases}
\label{eq:SM_phase}
\end{equation}
Thus $b_j^\dagger b_{j+2}$ has amplitude $Ke^{i\Phi}$ when its left endpoint is in the charger. Two such links cross the interface, starting at $j=N_C-1$ and $j=N_C$, when those links exist. All nearest-neighbor links and all next-nearest-neighbor links entirely inside the battery are real. This directed-link convention is the same in every method.

The internal battery Hamiltonian $H_B$ contains only terms acting entirely within the battery:
\begin{equation}
\begin{aligned}
H_B={}&\omega_B\sum_{j=N_C+1}^{N}n_j
+V\sum_{j=N_C+1}^{N-1}n_jn_{j+1}\\
&+J\sum_{j=N_C+1}^{N-1}(b_j^\dagger b_{j+1}+\mathrm{H.c.})\\
&+K\sum_{j=N_C+1}^{N-2}(b_j^\dagger b_{j+2}+\mathrm{H.c.}).
\end{aligned}
\label{eq:SM_HB}
\end{equation}
A hopping or interaction term crossing the interface belongs to $H(\Phi)$, but not to $H_B$. In particular, $H_B$ does not change with $\Phi$. The exact calculation, Gaussian reconstruction, and every occupation cutoff use this same operator.

The total number of excitations, $Q=\sum_j n_j$, is conserved. We work in the sector $Q=N_C$ and start with a filled charger and an empty battery:
\begin{equation}
|\psi(0)\rangle=
\Big(\prod_{j=1}^{N_C}b_j^\dagger\Big)|\mathrm{vac}\rangle.
\label{eq:SM_initial}
\end{equation}
Unless stated otherwise, $J=1$, $K=0.65J$, $\omega_C=J$, and $\omega_B=4J$. Interaction strength, charging time, and chain size are specified for each comparison. The full fixed-$Q$ basis contains $\binom NQ$ configurations; for $N_C=N_B=10$, this number is $184756$.

\suppnote{Exact evolution, reduced battery states, and unrestricted ergotropy}
\label{note:sm_exact}

We enumerate the occupation configurations with $Q$ excitations and construct the sparse Hermitian matrix of $H(\Phi)$. A hopping term connects two configurations only when one endpoint is occupied and the other is empty. Each matrix element uses the directed phase convention of Note~\ref{note:sm_hamiltonian}. The initial state in Eq.~\eqref{eq:SM_initial} then evolves as
\begin{equation}
|\psi(t,\Phi)\rangle=e^{-itH(\Phi)}|\psi(0)\rangle.
\label{eq:SM_exact_evolution}
\end{equation}
We apply the sparse exponential directly to the state; the independent verification solver uses \texttt{scipy.sparse.linalg.expm\_multiply}. We check Hermiticity and state normalization. The archived $N=20$ endpoint states are normalized to within accumulated floating-point propagation error.

The conserved total number makes the battery state block diagonal in its instantaneous occupation $q$. For each $q$, let $X_q$ contain the global-state amplitudes: rows label charger configurations with $Q-q$ excitations, and columns label battery configurations with $q$ excitations. Let $\rho_B^{(q)}$ be the unnormalized reduced block and $P_q$ its probability. Tracing out the charger gives
\begin{equation}
\rho_B=\bigoplus_q\rho_B^{(q)},\quad
\rho_B^{(q)}=X_q^{\mathsf T}X_q^*,\quad
P_q=\Tr\rho_B^{(q)}.
\label{eq:SM_blocks}
\end{equation}
For $P_q>0$, the normalized conditional state is $\sigma_q=\rho_B^{(q)}/P_q$, so $\rho_B=\bigoplus_qP_q\sigma_q$. The label $q$ does not count transfer events. An excitation can cross the interface and return repeatedly while the battery occupation stays below a fixed cutoff.

Write $H_B^{(q)}$ for the restriction of $H_B$ to occupation $q$. Its expectation can be evaluated block by block. Passive energy requires a different ordering: combine the eigenvalues of every $\rho_B^{(q)}$, including zeros, into a single descending list $p_a$, and combine the energies of the full $H_B$ into a single ascending list $\varepsilon_a$. The stored and passive energies are then
\begin{equation}
E_B=\sum_q\Tr\big[\rho_B^{(q)}H_B^{(q)}\big],\quad
E_{\mathrm{pass}}=\sum_{a=1}^{2^{N_B}}p_a\varepsilon_a.
\label{eq:SM_passive}
\end{equation}
The unrestricted ergotropy is $W_B=E_B-E_{\mathrm{pass}}$. Sorting separately within each $q$ block would instead restrict the extraction unitary to conserve battery excitation number. Note~\ref{note:sm_operational} studies that different quantity. It is not used in the unrestricted results.

Table~\ref{tab:SM_baseline} gives the exact energies at two fixed fluxes for $N=20$, $V=2J$, and $Jt=4$. Their ergotropy difference is $2.2078J$, favoring $\pi/2$. This comparison does not depend on locating a global flux maximum.

\begin{table}[htbp]
\caption{Exact stored, passive, and extractable energies at two fixed fluxes for $N_C=N_B=10$, $V=2J$, and $Jt=4$. Both columns use the unchanged full $H_B$; all energies are in units of $J$.}
\label{tab:SM_baseline}
\centering
\setlength{\tabcolsep}{15pt}
\begin{tabular}{lrr}
\hline\hline
Quantity & $\Phi=\pi$ & $\Phi=\pi/2$\\
\hline
$E_B/J$ & 8.1962 & 11.0143\\
$E_{\mathrm{pass}}/J$ & 2.6812 & 3.2914\\
$W_B/J$ & 5.5150 & 7.7228\\
\hline\hline
\end{tabular}
\end{table}

\suppnote{Jordan--Wigner transformation and Gaussian evolution}
\label{note:sm_gaussian}

A Jordan--Wigner transformation ordered from the left end of the chain maps the hard-core bosons to fermions $c_j$. Nearest-neighbor hopping is quadratic in these operators, but next-nearest-neighbor hopping retains the intermediate occupation:
\begin{equation}
\begin{aligned}
b_j^\dagger b_{j+1}&=c_j^\dagger c_{j+1},\\
b_j^\dagger b_{j+2}&=c_j^\dagger(1-2n_{j+1})c_{j+2}.
\end{aligned}
\label{eq:SM_JW}
\end{equation}
The term proportional to $K$ therefore contains occupation-dependent fermionic hopping even when $V=0$. By contrast, $K=V=0$ gives a quadratic Hamiltonian. The initial filled-charger state is a Slater determinant, and its exact evolution then remains Gaussian.

Define the one-body matrix by $C_{ij}=\langle c_j^\dagger c_i\rangle$. In a number-conserving Gaussian state, Wick's rule expresses higher-order averages in terms of this matrix. Applying it to the full Hamiltonian gives the Gaussian energy functional $\mathcal E_G(C)$:
\begin{equation}
\begin{aligned}
\mathcal E_G(C)&=\sum_j\omega_j C_{jj}
+V\sum_{j=1}^{N-1}\big(C_{jj}C_{j+1,j+1}-|C_{j,j+1}|^2\big)\\
&+2\operatorname{Re}\Bigg\{J\sum_{j=1}^{N-1}C_{j+1,j}+K\sum_{j=1}^{N-2}e^{i\phi_j}
\Big[C_{j+2,j}(1-2C_{j+1,j+1})+2C_{j+1,j}C_{j+2,j+1}\Big]\Bigg\}.
\end{aligned}
\label{eq:SM_gaussian_energy}
\end{equation}
The self-consistent one-body Hamiltonian $F_{\mathrm{HF}}(C)$ is defined by $\delta\mathcal E_G=\Tr[F_{\mathrm{HF}}(C)\,\delta C]$ for Hermitian variations $\delta C$. Equivalently, $(F_{\mathrm{HF}})_{ij}=\partial\mathcal E_G/\partial C_{ji}$ with a consistent matrix-index convention. The verification solver checks this relation by directional finite differences. Time-dependent Hartree--Fock (TDHF) evolves this matrix from the filled-charger initial condition:
\begin{equation}
\dot C_{\mathrm{HF}}=-i\big[F_{\mathrm{HF}}(C_{\mathrm{HF}}),C_{\mathrm{HF}}\big],\quad
C_{\mathrm{HF}}(0)=\operatorname{diag}(\underbrace{1,\ldots,1}_{N_C},\underbrace{0,\ldots,0}_{N_B}).
\label{eq:SM_TDHF}
\end{equation}
TDHF retains a Gaussian state throughout the evolution. The implementation includes both the interaction $V$ and the occupation-dependent terms generated by $K$; it does not replace hard-core next-nearest-neighbor hopping by a purely quadratic term.

As a check, exact propagation, TDHF, and Gaussian reconstruction from exact one-body data agree to numerical precision at $K=V=0$. Setting only $V=0$ does not give this agreement when $K=0.65J$. For example, at $N=16$ and $Jt=4$, the exact ergotropy at $\pi/2$ minus that at $\pi$ is $-1.018923J$, whereas TDHF gives $-1.217530J$. The occupation-dependent factor in Eq.~\eqref{eq:SM_JW} remains present in this case.

\suppnote{Correlations missing from the one-body equation}
\label{note:sm_source}

The exact equation for $C$ follows from the full Hamiltonian. Its quadratic terms and the Wick-factorized part of its four-fermion averages give the TDHF commutator. The remaining part depends on correlations not fixed by $C$. One such connected four-fermion average is
\begin{equation}
\begin{aligned}
\mathcal C_{ijkl}=\langle c_i^\dagger c_j^\dagger c_kc_l\rangle
-\langle c_i^\dagger c_l\rangle\langle c_j^\dagger c_k\rangle+\langle c_i^\dagger c_k\rangle\langle c_j^\dagger c_l\rangle.
\end{aligned}
\label{eq:SM_cumulant}
\end{equation}
These quantities vanish in a number-conserving Gaussian state. Their weighted contribution to the exact one-body equation is $\mathcal I_{\mathrm{corr}}$ in Eq.~\eqref{eq:correlation_source}. We evaluate it directly from the state derivative $\dot\psi=-iH\psi$, rather than from a finite difference in time:
\begin{equation}
\mathcal I_{\mathrm{corr}}
=\dot C_{\mathrm{exact}}+i\big[F_{\mathrm{HF}}(C_{\mathrm{exact}}),C_{\mathrm{exact}}\big],
\quad \dot\psi=-iH\psi.
\label{eq:SM_source}
\end{equation}
Contracting $\dot\psi$ and $\psi$ with the one-body operators gives $\dot C_{\mathrm{exact}}$. We check that $C$ is Hermitian, that $\Tr\mathcal I_{\mathrm{corr}}=0$, and that the source vanishes in the quadratic limit. It is also zero in the initial filled-charger Slater determinant. It becomes nonzero as the interacting evolution builds correlations.

Table~\ref{tab:SM_source} compares the source and the exact--TDHF one-body difference at a common flux and time. The symbol $\|\cdot\|_F$ denotes the Frobenius matrix norm. Figure~\figpanel{fig:SM_diagnostics}{a} shows the corresponding time dependence. The source is nonzero already at $V=0$, $K=0.65J$, and is larger at the final time for the listed interacting cases. A larger source need not imply a particular sign of the work slope or a particular optimum: both the full trajectory and the battery spectrum matter.

\begin{table}[htbp]
\caption{Connected-source norm and exact--TDHF one-body difference at $N_C=N_B=8$, $Jt=4$, and $\Phi=0.76\pi$. The superscripts ex and HF denote exact and TDHF evolution. The Frobenius source norm has inverse-time units, equal to energy units because $\hbar=1$.}
\label{tab:SM_source}
\centering
\setlength{\tabcolsep}{15pt}
\begin{tabular}{cccc}
\hline\hline
$K/J$ & $V/J$ & $\|\mathcal I_{\mathrm{corr}}\|_F/J$ & $\|C_{\mathrm{ex}}-C_{\mathrm{HF}}\|_F$\\
\hline
0 & 0 & $<3\times10^{-14}$ & $2.4\times10^{-9}$\\
0.65 & 0 & 0.368221 & 0.371627\\
0.65 & 1.5 & 0.693751 & 1.059357\\
0.65 & 2 & 0.765643 & 1.518856\\
\hline\hline
\end{tabular}
\end{table}

Equation~\eqref{eq:correlation_source} is an exact relation in the hierarchy of correlation equations. It does not close the evolution of $C$ unless the missing higher-order information is also supplied.

\suppnote{Gaussian reconstruction, flux selection, and work response}
\label{note:sm_reconstruction}

\subsection{Reconstructing a Gaussian state}

Let $C_B$ be the battery submatrix of $C$. Diagonalizing it gives occupation eigenvalues $0\le\nu_\ell\le1$, one for each battery fermionic mode. Here a number-conserving Gaussian state commutes with excitation number, but need not have a fixed excitation number. The reconstructed state $\mathcal G(C_B)$ has independent occupations $n_\ell\in\{0,1\}$ of these modes, so its eigenvalues are
\begin{equation}
p_{\bm n}^{G}=\prod_{\ell=1}^{N_B}
\nu_\ell^{n_\ell}(1-\nu_\ell)^{1-n_\ell},\quad
n_\ell\in\{0,1\}.
\label{eq:SM_G_weights}
\end{equation}
The weights sum to one. We calculate $E_G(C_B)=\Tr[\mathcal G(C_B)H_B]$ by Wick's rule and obtain its passive energy by pairing the descending product weights with the ascending energies of the full $H_B$. Their difference is $W_G(C_B)$. The exact state and its Gaussian reconstruction share $C_B$, but generally differ in occupation-sector probabilities and higher-order correlations. The reconstruction is therefore a comparison state, not an independently evolved charging state.

Adding and subtracting $W_G(C_B^{\mathrm{ex}})$ gives Eq.~\eqref{eq:gaussian_ledger}. Its first bracket changes the one-body input within the Gaussian description; its second changes the state while keeping that input exact. Both depend on flux and contribute to the same ergotropy difference. They are not separately available amounts of work.

For any method $M$, let $\Phi_M^*$ be its best phase found in the numerical search and let $\Phi_{\mathrm{ex}}^*$ be the exact reference. We assess the chosen control through its exact ergotropy regret, $\mathcal R_M=W_B^{\mathrm{ex}}(\Phi_{\mathrm{ex}}^*)-W_B^{\mathrm{ex}}(\Phi_M^*)$. Both terms are evaluated by exact propagation. This separates the quality of the control from the accuracy of the method's own ergotropy values.

\subsection{Comparing maxima across the full period}

We call $0\le\Phi<\pi$ the positive branch and $\pi\le\Phi<2\pi$ the negative branch; the latter is equivalent to $-\pi\le\Phi<0$. These labels specify flux ranges, not a separately measured direction of transport. Table~\ref{tab:SM_oracle_opt} and Figs.~\figpanel{fig:SM_diagnostics}{b} and \figpanel{fig:SM_diagnostics}{d} compare exact, TDHF, and exact-data Gaussian selections. The table also retains the Gaussian positive-branch local maximum $G_+$, which need not be its highest maximum over the full period, $G_{\mathrm{per}}$.

\begin{table}[htbp]
\caption{Flux choices and exact regrets for $N_C=N_B=8$, $K=0.65J$, and $Jt=4$. Phases are divided by $\pi$. $G_+$ is the positive-branch maximum of the Gaussian reconstructed from exact $C_B$; $G_{\mathrm{per}}$ is its highest maximum found over $[0,2\pi)$. The last two columns evaluate the exact work lost at those choices.}
\label{tab:SM_oracle_opt}
\centering
\setlength{\tabcolsep}{15pt}
\begin{tabular}{rrrrrrr}
\hline\hline
$V/J$ & Exact & TDHF & $G_+$ & $G_{\mathrm{per}}$ & $\mathcal R(G_+)/J$ & $\mathcal R(G_{\mathrm{per}})/J$\\
\hline
1.5 & 0.634376 & 0.811802 & 0.589097 & 0.589097 & 0.013087 & 0.013087\\
2 & 0.465775 & 0.759651 & 0.454541 & 1.487602 & 0.004247 & 0.748253\\
2.5 & 0.388618 & 0.719109 & 0.369383 & 1.511760 & 0.013750 & 0.623824\\
\hline\hline
\end{tabular}

\end{table}

At $V=1.5J$, the reconstructed Gaussian selects the positive branch. At $V=2J$ and $2.5J$, its higher maximum lies near $3\pi/2$, despite a positive-branch local peak close to the exact choice. The small regrets $0.004247J$ and $0.013750J$ belong only to those local positive-branch choices. Selecting the Gaussian's higher full-period maximum instead gives regrets $0.748253J$ and $0.623824J$. Table~\ref{tab:SM_gaussian_branches} shows directly that the Gaussian reconstruction and the exact battery state rank these two phases in opposite order.

\begin{table}[htbp]
\caption{Gaussian and exact ergotropies, in units of $J$, at the two Gaussian selections in Table~\ref{tab:SM_oracle_opt}. At $V=2J$ and $2.5J$, the exact state favors the positive-branch choice while the reconstruction favors its full-period choice. At $V=1.5J$, the choices coincide.}
\label{tab:SM_gaussian_branches}
\centering
\setlength{\tabcolsep}{15pt}
\begin{tabular}{rrrrr}
\hline\hline
$V/J$ & $W_G(\Phi_{G_+})/J$ & $W_G(\Phi_{G_{\mathrm{per}}})/J$ & $W_B(\Phi_{G_+})/J$ & $W_B(\Phi_{G_{\mathrm{per}}})/J$\\
\hline
1.5 & 4.816819 & 4.816819 & 7.169528 & 7.169528\\
2 & 4.181392 & 4.478504 & 7.242165 & 6.498159\\
2.5 & 3.556826 & 4.254659 & 7.302378 & 6.692304\\
\hline\hline
\end{tabular}

\end{table}

An additional archived comparison at $V=2J$, $Jt=5$ gives positive-branch choices $0.460039\pi$ for exact evolution, $0.765791\pi$ for TDHF, and $0.424463\pi$ for Gaussian reconstruction. The latter two have exact regrets $1.427644J$ and $0.034343J$ relative to that reference. These local choices are retained separately from the later full-period interaction--time scan below.

\subsection{Local response near a TDHF maximum}

\begin{figure}[htbp]
\centering
\includegraphics[width=0.80\linewidth]{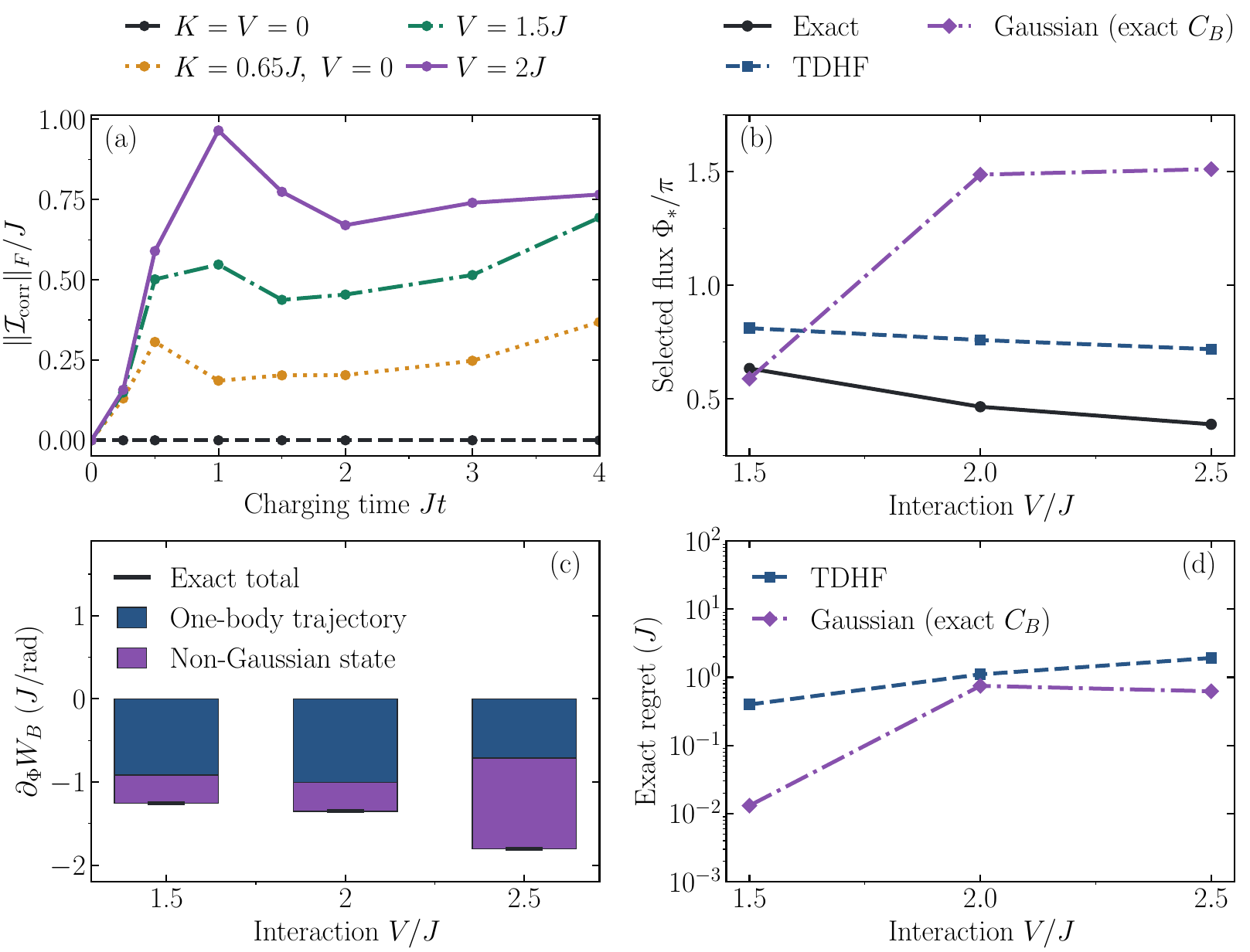}
\caption{Gaussian diagnostics at $N_C=N_B=8$. (a) Time dependence of the connected-source Frobenius norm at $\Phi=0.76\pi$ for the indicated couplings. (b) Highest-peak flux found over the full period for exact evolution, TDHF, and Gaussian reconstruction from exact $C_B$. (c) One-body and non-Gaussian contributions to the exact slope at the TDHF maximum; black marks give their sum. (d) Exact ergotropy regrets of the choices in (b). Here $K=0.65J$ except in the quadratic control, and $Jt=4$ in (b)--(d). Lines join sampled values; they do not locate branch transitions.}
\label{fig:SM_diagnostics}
\end{figure}

To understand how the work changes near a particular phase, let $\Phi_{\mathrm{HF}}$ be a smooth TDHF maximum. The derivative of its own Gaussian ergotropy vanishes there. Denote the derivatives of the two brackets in Eq.~\eqref{eq:gaussian_ledger} by $g_C$ and $g_{\mathrm{NG}}$, respectively. They then give the exact slope:
\begin{equation}
\begin{aligned}
\left.\partial_\Phi W_B\right|_{\Phi_{\mathrm{HF}}}
={}&\underbrace{\left.\partial_\Phi
\big[W_G(C_B^{\mathrm{ex}})-W_G(C_B^{\mathrm{HF}})\big]\right|_{\Phi_{\mathrm{HF}}}}_{g_C}\\
&+\underbrace{\left.\partial_\Phi
\big[W_B-W_G(C_B^{\mathrm{ex}})\big]\right|_{\Phi_{\mathrm{HF}}}}_{g_{\mathrm{NG}}}.
\end{aligned}
\label{eq:SM_torque}
\end{equation}
We check the derivatives with phase steps of $0.002$ and $0.001$ radians. Table~\ref{tab:SM_torque} gives the results, also shown in Fig.~\figpanel{fig:SM_diagnostics}{c}. The one-body correction is larger at $V=2J$, whereas the non-Gaussian state correction is larger at $V=2.5J$. Neither contribution always dominates.

\begin{table}[htbp]
\caption{Local ergotropy slopes at the TDHF-selected phase for $N_C=N_B=8$ and $Jt=4$, in $J$ per radian. The one-body and non-Gaussian contributions sum to the exact slope, up to rounding.}
\label{tab:SM_torque}
\centering
\setlength{\tabcolsep}{15pt}
\begin{tabular}{rrrr}
\hline\hline
$V/J$ & $g_C$ & $g_{\mathrm{NG}}$ & $\partial_\Phi W_B$\\
\hline
1.5 & $-0.917073$ & $-0.338532$ & $-1.255604$\\
2.0 & $-1.003027$ & $-0.348094$ & $-1.351120$\\
2.5 & $-0.711093$ & $-1.091876$ & $-1.802968$\\
\hline\hline
\end{tabular}
\end{table}

This is a local comparison at one phase. It does not determine the relative heights of distant maxima. A large one-body contribution to the local slope therefore does not establish that exact one-body data recover the full-period choice.

Both slope terms can be separated further into stored- and passive-energy parts. Write $E_{\mathrm{pass},G}(C_B)=E_G(C_B)-W_G(C_B)$. At the same phase $\Phi_{\mathrm{HF}}$, define the stored-energy contributions $g_C^E,g_{\mathrm{NG}}^E$ and passive-energy contributions $g_C^P,g_{\mathrm{NG}}^P$ by
\begin{equation}
\begin{aligned}
g_C^E&=\partial_\Phi[E_G(C_B^{\mathrm{ex}})-E_G(C_B^{\mathrm{HF}})],
&g_C^P&=\partial_\Phi[E_{\mathrm{pass},G}(C_B^{\mathrm{ex}})-E_{\mathrm{pass},G}(C_B^{\mathrm{HF}})],\\
g_{\mathrm{NG}}^E&=\partial_\Phi[E_B-E_G(C_B^{\mathrm{ex}})],
&g_{\mathrm{NG}}^P&=\partial_\Phi[E_{\mathrm{pass}}-E_{\mathrm{pass},G}(C_B^{\mathrm{ex}})].
\end{aligned}
\label{eq:SM_energy_passive_split}
\end{equation}
Thus $g_C=g_C^E-g_C^P$ and $g_{\mathrm{NG}}=g_{\mathrm{NG}}^E-g_{\mathrm{NG}}^P$. Table~\ref{tab:SM_energy_passive_split} shows that the one-body correction mainly changes the stored-energy slope in this benchmark. The non-Gaussian correction mainly changes the passive-energy slope. These are parts of the same work response, not independent resources.

\begin{table}[htbp]
\caption{Stored- and passive-energy contributions at the TDHF maximum for $N_C=N_B=8$, $V=2J$, and $Jt=4$. Values are in $J$ per radian and use a central difference with step $0.0005$ rad. The final column is the stored contribution minus the passive contribution.}
\label{tab:SM_energy_passive_split}
\centering
\setlength{\tabcolsep}{15pt}
\begin{tabular}{lrrr}
\hline\hline
Contribution & $g^E$ & $g^P$ & $g^E-g^P$\\
\hline
One-body trajectory & $-1.140192$ & $-0.137166$ & $-1.003026$\\
Non-Gaussian state & $0.018824$ & $0.366919$ & $-0.348094$\\
\hline\hline
\end{tabular}

\end{table}

For comparison with the fixed-flux result in Fig.~\ref{fig:flux_failure}, at $N=20$ the exact-data Gaussian gives $W_G(\pi/2)-W_G(\pi)=1.529511423J$, while the exact difference is $2.207838287J$. TDHF gives the opposite sign. This checks the ordering of two supplied phases; it is not a search for the reconstructed-Gaussian optimum at $N=20$.

\subsection{Full-period comparison with the reduced many-body model}

Figure~\ref{fig:fullperiod} of the Letter compares three entire ergotropy curves: exact evolution, Gaussian reconstruction from the exact $C_B$ at every flux, and independent evolution of the correlated model retaining $q\le3$. The last method is the occupation-sector CI projection of Eq.~\eqref{eq:ci}, described in Note~\ref{note:sm_ci}. Every method uses unrestricted extraction and the same complete, flux-independent $H_B$. Only the exact-data Gaussian needs the full exact trajectory as input.

For method $M$, write $W_{M,+}^*=\max_{0\le\Phi<\pi}W_M(\Phi)$, $W_{M,-}^*=\max_{\pi\le\Phi<2\pi}W_M(\Phi)$, and $\Delta_{+-}^M=W_{M,+}^*-W_{M,-}^*$. Positive $\Delta_{+-}^M$ favors the positive branch. At the tested points the selected branch peaks lie inside their intervals.

At $N=16$, $V=2J$, and $Jt=4$, the refined positive- and negative-branch phases are $(0.465775,1.476098)\pi$ for exact evolution, $(0.454541,1.487602)\pi$ for the exact-data Gaussian, and $(0.473872,1.475886)\pi$ for $m=3$. Their branch advantages are $+0.744109J$, $-0.297112J$, and $+0.707677J$, respectively [Fig.~\figpanel{fig:fullperiod}{b}]. The Gaussian-selected phase loses $0.748253J$ of exact ergotropy; the $m=3$ choice loses $0.001856J$. The reduced model uses $3985$ of $12870$ configurations, about $31\%$, and recovers the branch ranking without reproducing the ergotropy exactly.

To see which energy contributions change this ranking, Table~\ref{tab:SM_fullperiod_pair} uses the same two phases for every method: the Gaussian local maxima $\Phi_+=0.454540645\pi$ and $\Phi_-=1.487602343\pi$. Define $\Delta E=E(\Phi_+)-E(\Phi_-)$ and similarly $\Delta E_{\mathrm{pass}}$ and $\Delta W$. These fixed-pair differences are not the differences between each method's own maxima in Fig.~\figpanel{fig:fullperiod}{b}.

\begin{table}[htbp]
\caption{Energy differences, in units of $J$, between the common pair $\Phi_+=0.454540645\pi$ and $\Phi_-=1.487602343\pi$ at $N=16$, $V=2J$, and $Jt=4$. The Gaussian state uses exact one-body data at each phase. The last column is $\Delta W=\Delta E-\Delta E_{\mathrm{pass}}$.}
\label{tab:SM_fullperiod_pair}
\centering
\setlength{\tabcolsep}{15pt}
\begin{tabular}{lrrr}
\hline\hline
Method & $\Delta E/J$ & $\Delta E_{\mathrm{pass}}/J$ & $\Delta W/J$\\
\hline
Exact many-body & $2.197688$ & $1.453682$ & $+0.744006$\\
Gaussian of exact $C_B$ & $1.853082$ & $2.150194$ & $-0.297112$\\
Correlated $m=3$ & $1.811149$ & $1.110517$ & $+0.700632$\\
\hline\hline
\end{tabular}
\end{table}

The Gaussian reconstruction makes the passive-energy increase larger than the stored-energy increase, so it favors the negative-branch phase. Exact evolution and correlated $m=3$ give the opposite balance. Both stored and passive energy therefore contribute to the difference between the reconstructed and physical states.

The smaller cutoff $q\le2$ retains $849$ of $12870$ configurations, about $6.6\%$. It selects $0.468199\pi$ and has exact regret $1.9028\times10^{-4}J$, even smaller than for $m=3$ at this point. Its own branch advantage is only $0.241201J$, so it reproduces the exact curve less accurately while choosing a better control. The next comparison tests why this single-point success does not establish that the smaller cutoff is always sufficient.

\subsection{Robustness across interaction strength and charging time}

We keep $N_C=N_B=8$, $K=0.65J$, $\omega_C=J$, and $\omega_B=4J$, and test nine equally spaced values $1.5\le V/J\le2.5$ and seven times $3.5\le Jt\le5$. At each of the 63 combinations, we evaluate exact evolution, Gaussian reconstruction from exact $C_B(\Phi)$, and the independently evolved $m=2,3$ models on 128 equally spaced fluxes over $[0,2\pi)$. We then refine the sampled maxima by direct evaluation of the relevant objective. We do not infer peak locations from interpolated heatmap colors.

\begin{figure}[htbp]
\centering
\includegraphics[width=0.8\linewidth]{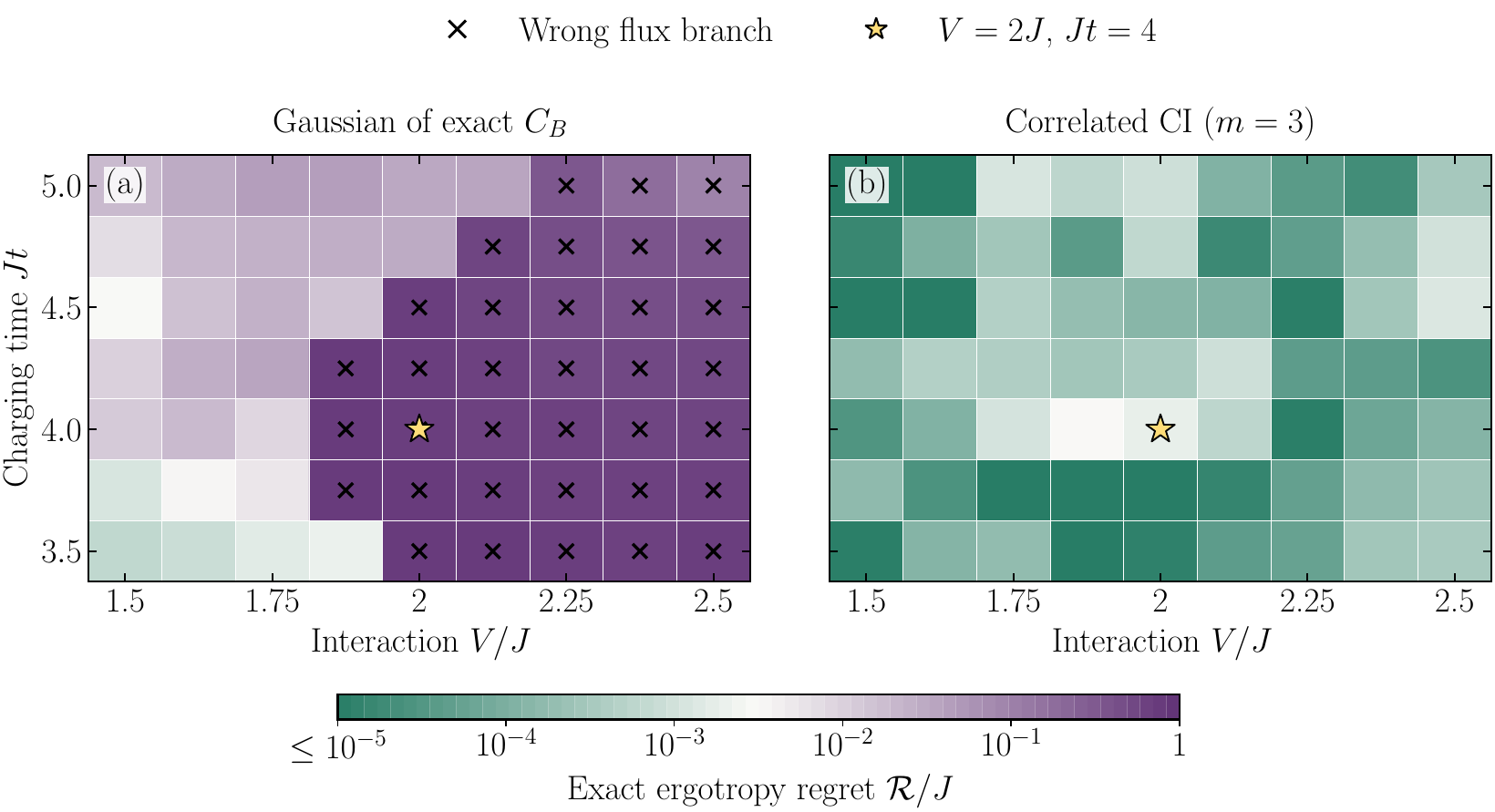}
\caption{Exact ergotropy lost by choosing the flux selected by (a) Gaussian reconstruction from exact $C_B(\Phi)$ and (b) independently evolved correlated CI with $m=3$. Both panels use the same logarithmic scale for $\mathcal R/J$; values below $10^{-5}$ share the lowest color. Black crosses mark choices on the opposite branch to exact dynamics. The star marks $V=2J$, $Jt=4$. Here $N_C=N_B=8$, $K=0.65J$, $\omega_C=J$, and $\omega_B=4J$. The $9\times7$ cells show computed parameter points, not a guarantee for all points between them.}
\label{fig:SM_robustness_regret}
\end{figure}

For every selected phase we compute the regret $\mathcal R_M$ with exact many-body evolution. We also compare the sign of each method's branch advantage $\Delta_{+-}^M$ with the exact sign. Figures~\figpanel{fig:SM_robustness_regret}{a} and \figpanel{fig:SM_robustness_regret}{b} show the regrets of the exact-data Gaussian and $m=3$ model on a common color scale. Black crosses identify wrong-branch choices. Table~\ref{tab:SM_robustness} gives the counts and errors, including the smaller cutoff.

\begin{table}[htbp]
\caption{Full-period flux selection on the 63-point grid in Fig.~\ref{fig:SM_robustness_regret}. Wrong-branch counts compare the signs of the separately optimized branch advantages. All regrets use exact evolution at the selected phases; the final two columns give their median and largest values.}
\label{tab:SM_robustness}
\centering
\setlength{\tabcolsep}{15pt}
\begin{tabular}{lrrr}
\hline\hline
Selector & Wrong branch / 63 & Median $\mathcal R/J$ & Max. $\mathcal R/J$\\
\hline
Gaussian of exact $C_B$ & 35 & $0.3519$ & $0.8227$\\
Correlated $m=3$ & 0 & $1.23\times10^{-4}$ & $3.43\times10^{-3}$\\
Correlated $m=2$ & 5 & $2.42\times10^{-3}$ & $0.9782$\\
\hline\hline
\end{tabular}
\end{table}

Exact dynamics favors the positive branch at every grid point. The exact-data Gaussian instead chooses the negative branch at 35 points. The $m=3$ model agrees with the exact branch everywhere and has regret at most $3.43\times10^{-3}J$, compared with $0.8227J$ for the Gaussian. Thus the Gaussian failure is not confined to the reference point. This is evidence within a finite, retrospectively chosen window, not a guarantee outside it.

\begin{figure}[htbp]
\centering
\includegraphics[width=0.9\linewidth]{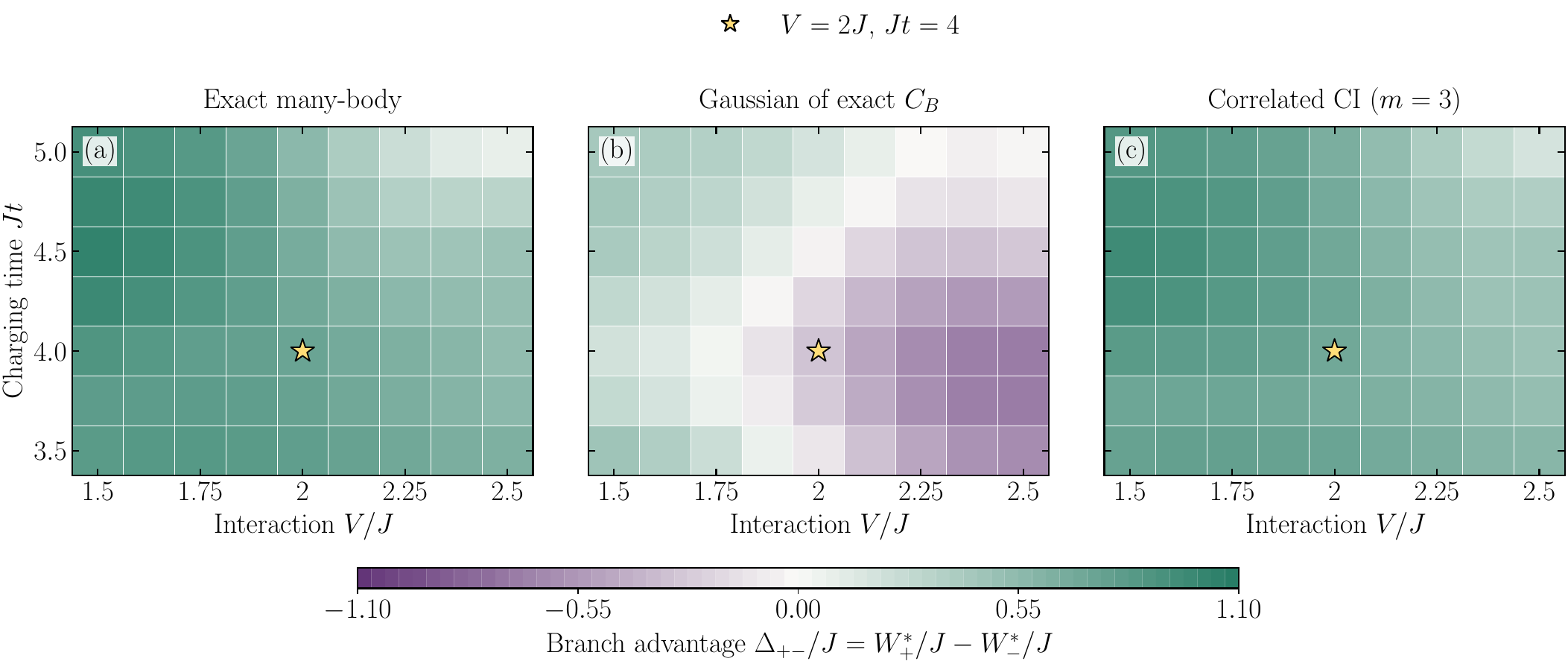}
\caption{Positive-minus-negative branch advantage $\Delta_{+-}^M/J$ for (a) exact dynamics, (b) Gaussian reconstruction from exact $C_B$, and (c) correlated $m=3$. Green values favor $0\le\Phi<\pi$; purple values favor $\pi\le\Phi<2\pi$. Each value compares separately optimized peaks. All panels use the same scale, grid, and physical parameters as Fig.~\ref{fig:SM_robustness_regret}; the star marks the reference point.}
\label{fig:SM_robustness_branches}
\end{figure}

Figure~\ref{fig:SM_robustness_branches} shows the branch preference itself: panel (a) is exact, panel (b) uses the exact-data Gaussian, and panel (c) uses correlated $m=3$. The change of sign in panel (b), absent from (a) and (c), identifies the region of incorrect Gaussian ranking. Since exact dynamics always favors the positive branch here, branch agreement alone cannot establish accurate control. A fixed $\Phi=\pi/2$ has median exact regret $0.1116J$, and $m=3$ does better at 62 of 63 points. It therefore improves the selected phase within that branch as well.

\begin{figure}[htbp]
\centering
\includegraphics[width=0.8\linewidth]{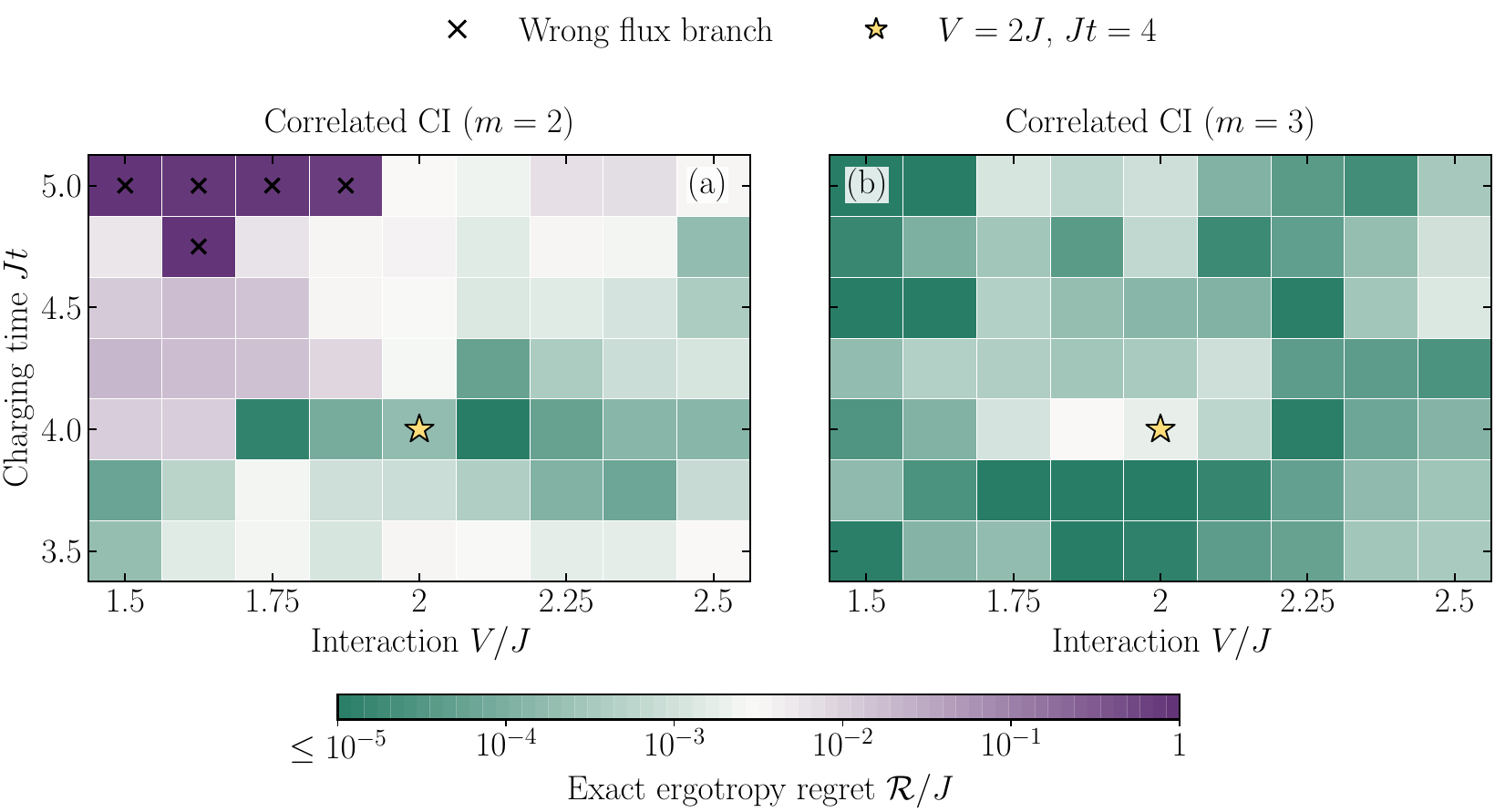}
\caption{Effect of the occupation cutoff on exact ergotropy regret: (a) $m=2$ and (b) $m=3$ on the same 63-point grid. Both use the same logarithmic scale. Crosses mark wrong-branch choices, and the star marks $V=2J$, $Jt=4$. The smaller cutoff is accurate at the reference point but fails the branch test at five sampled points. Values below $10^{-5}J$ share a display color; the data retain their unmodified values.}
\label{fig:SM_cutoff_robustness}
\end{figure}

The smaller cutoff does not keep this accuracy throughout the window. Figure~\figpanel{fig:SM_cutoff_robustness}{a} shows that $m=2$ selects the wrong branch at five points, while panel (b) repeats the $m=3$ result for comparison. At $V=1.5J$, $Jt=5$, the $m=2$ regret is about $0.953J$, whereas the $m=3$ regret is below $10^{-5}J$. This failure shows why the good $m=2$ choice at the original benchmark cannot be taken as a general cutoff rule.

These 63 points form a structured retrospective scan, not independent prospective tests. Local maxima are refined from finite flux grids, which cannot exclude every possible unsampled narrow peak. State-norm and independent propagation checks are included with the data. The figures establish behavior at the sampled parameters, not certified boundaries between continuous regions.

\subsection{Dependence on the phase-bearing hopping $K$}

The strength $K$ controls both triangular interference and the occupation-dependent hopping term in Eq.~\eqref{eq:SM_JW}. We therefore vary $K/J$ at fixed $N_C=N_B=8$, $V=2J$, $Jt=4$, $\omega_C=J$, and $\omega_B=4J$.
For each value of $K$, all methods use the same complete battery Hamiltonian $H_B$, which changes with $K$ but remains independent of $\Phi$.
Hence the comparison at a given $K$ isolates the quality of flux selection; it does not compare the absolute stored work of different battery Hamiltonians. At $K=0$ the Hamiltonian is independent of $\Phi$ and there is no unique optimal flux, so this control case is excluded from branch statistics.

\begin{figure}[htbp]
\centering
\includegraphics[width=0.9\linewidth]{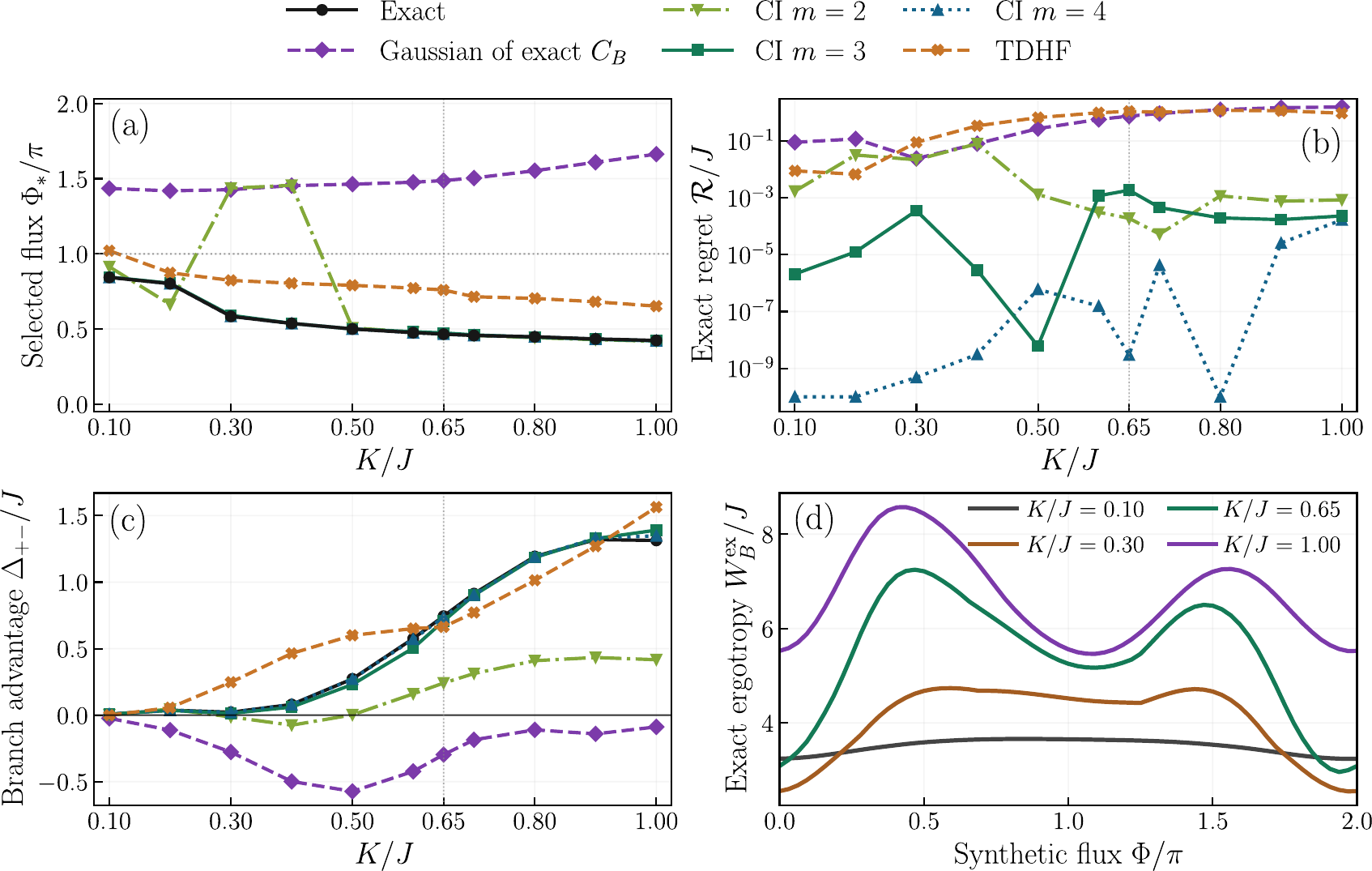}
\caption{Sensitivity to the phase-bearing hopping amplitude $K$ at $N_C=N_B=8$, $V=2J$, and $Jt=4$.
(a) Selected optimal fluxes over $[0,2\pi)$ for exact evolution, Gaussian reconstruction from exact $C_B$, correlated cutoffs $m=2,3,4$, and TDHF.
(b) Exact ergotropy regret at each method's selected flux; values below $10^{-10}J$ are displayed at the logarithmic plotting floor.
(c) Positive-minus-negative branch advantage of each method; the
horizontal zero line separates the preferred branches.
(d) Exact ergotropy curves for four representative values of $K/J$.
Vertical dotted lines in (a)--(c) mark the original $K/J=0.65$ benchmark. Lines connect sampled results and do not establish behavior between them. The complete $H_B$ is held fixed against $\Phi$ for each $K$, but its spectrum can change when $K$ changes.}
\label{fig:SM_K_robustness}
\end{figure}

For eleven nonzero values $K/J=0.1,0.2,\ldots,1.0$, with the additional reference point $K/J=0.65$, we search the full interval $0\le\Phi<2\pi$. Figure~\ref{fig:SM_K_robustness} and Table~\ref{tab:SM_K_robustness} show the selected phases, exact regrets, and differences between branch maxima.
Exact evolution favors the positive branch at all eleven sampled points. Gaussian reconstruction from exact one-body data instead favors the negative branch at every point; its largest exact regret is $1.5674J$. The $m=2$ model selects the wrong branch at $K/J=0.3$ and $0.4$. By contrast, the independently evolved $m=3$ and $m=4$ models select the correct branch at all eleven points, with respective largest exact regrets $1.856\times10^{-3}J$ and $1.679\times10^{-4}J$.
Both also improve on the fixed choice $\Phi=\pi/2$ at every sampled nonzero $K$. TDHF selects the wrong branch only at $K/J=0.1$, but its largest exact regret reaches $1.1923J$: correctly identifying the branch does not necessarily locate a good control within it.

\begin{table}[htbp]
\caption{Full-period flux selection with varying hopping $K/J$.
For each $K$, the exact phase $\Phi_*^{\mathrm{ex}}$ is shown,
and the remaining columns give the exact unrestricted-ergotropy regret at the flux selected by each approximate method, in units of $J$. Gaussian reconstruction uses exact one-body data. It selects the wrong branch throughout; CI $m=2$ selects the wrong branch at $K/J=0.3$ and $0.4$. Entries below $10^{-8}J$ are rounded to a numerical floor, not claimed as certified accuracy. All maxima are numerical full-period searches.}
\label{tab:SM_K_robustness}
\centering
\setlength{\tabcolsep}{8pt}
\begin{tabular}{rrrrrr}
\hline\hline
$K/J$ & $\Phi_*^{\mathrm{ex}}/\pi$ & Gaussian & CI $m=2$ & CI $m=3$ & CI $m=4$\\
\hline
$0.10$ & $0.843300$ & $0.0900$ & $1.72\times10^{-3}$ & $2.09\times10^{-6}$ & $<10^{-8}$ \\
$0.20$ & $0.801713$ & $0.1185$ & $0.0321$ & $1.26\times10^{-5}$ & $<10^{-8}$ \\
$0.30$ & $0.584196$ & $0.0241$ & $0.0219$ & $3.55\times10^{-4}$ & $<10^{-8}$ \\
$0.40$ & $0.536959$ & $0.0805$ & $0.0801$ & $2.86\times10^{-6}$ & $<10^{-8}$ \\
$0.50$ & $0.500292$ & $0.2732$ & $1.31\times10^{-3}$ & $<10^{-8}$ & $6.07\times10^{-7}$ \\
$0.60$ & $0.475752$ & $0.5766$ & $3.12\times10^{-4}$ & $1.19\times10^{-3}$ & $1.55\times10^{-7}$ \\
$0.65$ & $0.465775$ & $0.7483$ & $1.90\times10^{-4}$ & $1.86\times10^{-3}$ & $<10^{-8}$ \\
$0.70$ & $0.456932$ & $0.9270$ & $5.31\times10^{-5}$ & $4.56\times10^{-4}$ & $4.27\times10^{-6}$ \\
$0.80$ & $0.448226$ & $1.2640$ & $1.17\times10^{-3}$ & $1.97\times10^{-4}$ & $<10^{-8}$ \\
$0.90$ & $0.433898$ & $1.4924$ & $7.64\times10^{-4}$ & $1.71\times10^{-4}$ & $2.58\times10^{-5}$ \\
$1.00$ & $0.424027$ & $1.5674$ & $8.50\times10^{-4}$ & $2.31\times10^{-4}$ & $1.68\times10^{-4}$ \\
\hline\hline
\end{tabular}
\end{table}

The numerical searches sample 64 equally spaced flux values, refine the sampled local peaks within each branch by direct propagation, and compare the resulting branch maxima. The independent solver reproduces the original $K/J=0.65$ benchmark and checks Hermiticity, state normalization, total-energy conservation, and the quadratic $K=V=0$ limit. Additional 128-point flux-grid checks at $K/J=0.1,0.3,0.65,0.9$ confirm both the selected branches and the reported regrets within numerical precision. The scan is controlled in $K$ but finite and sampled: it does not certify continuous global optima, transitions between grid points, or a universal occupation cutoff.

\subsection{Occupation-resolved work response}

The block structure also gives an exact response identity that does not rely on triangular geometry. Write $\rho_B=\bigoplus_q\rho_B^{(q)}$, $H_B=\bigoplus_qH_B^{(q)}$, and $\rho_B^{(q)}=P_q\sigma_q$. Let $p_{qa}$ and $|r_{qa}\rangle$ be the eigenvalues and eigenvectors of each unnormalized state block. Order all $p_{qa}$ together in descending order, with global rank $s(q,a)$. If $\varepsilon_1\le\varepsilon_2\le\cdots$ are the energies of the full $H_B$, define the passive-assignment operator $\mathcal K_{\rho_B}$ by
\begin{equation}
\begin{aligned}
E_{\mathrm{pass}}(\rho_B)
&=\sum_{q,a}\varepsilon_{s(q,a)}p_{qa},\\
\mathcal K_{\rho_B}
&=\sum_{q,a}\varepsilon_{s(q,a)}|r_{qa}\rangle\langle r_{qa}|.
\end{aligned}
\label{eq:SM_passive_assignment}
\end{equation}
The state eigenvectors can be chosen within occupation sectors, so $\mathcal K_{\rho_B}$ is block diagonal. Its assigned energies, however, come from the full globally sorted Hamiltonian spectrum and need not belong to the same $q$ sector as their assigned eigenvectors. This construction does not restrict extraction to number-conserving unitaries.

Where the passive energy is differentiable, differentiating the ordered eigenvalue sum gives $\partial_\Phi E_{\mathrm{pass}}=\Tr[\mathcal K_{\rho_B}\,\partial_\Phi\rho_B]$. No derivative of $\mathcal K_{\rho_B}$ is needed: the expression already differentiates the state eigenvalues with their assigned energies. Since $H_B$ is fixed, the ergotropy response is
\begin{equation}
\partial_\Phi W_B
=\Tr\big[(H_B-\mathcal K_{\rho_B})\partial_\Phi\rho_B\big].
\label{eq:SM_general_response}
\end{equation}
For $P_q>0$, use $\partial_\Phi\rho_B^{(q)}=(\partial_\Phi P_q)\sigma_q+P_q\partial_\Phi\sigma_q$. Taking the trace sector by sector, with $\mathcal K_q$ the $q$ block of $\mathcal K_{\rho_B}$, gives Eq.~\eqref{eq:sector_response}. Its two terms describe changes in occupation probabilities and changes in the conditional states. They add to one derivative; they are neither independent sources of work nor a closed rule for the optimum.

For an unoccupied sector, the form $\partial_\Phi W_B=\sum_q\Tr[(H_B^{(q)}-\mathcal K_q)\partial_\Phi\rho_B^{(q)}]$ avoids division by $P_q$. At an isolated crossing of globally ranked state eigenvalues, passive energy can have a kink. The response must then be evaluated from each side with the appropriate ordering. Persistent zero eigenvalues do not by themselves invalidate the expression along a smooth path of fixed rank.

The archived tangent-state calculation at $N=20$, $V=2J$, $Jt=4$, and $\Phi=0.4689\pi$ gives $-0.0141J$ per radian from occupation probabilities and $+0.0142J$ per radian from conditional states. Their sum, $+0.00008J$ per radian, agrees with the stored-energy slope $+0.0175J$ minus the passive-energy slope $+0.0174J$, both per radian. The near cancellation illustrates why occupation probabilities alone cannot locate this optimum. Evaluating the identity still requires the flux-dependent state.

\suppnote{Changing occupation-sector couplings and passive-energy ordering}
\label{note:sm_transfer}

At fixed $Q=N_C$, let $\Pi_q$ project onto configurations with exactly $q$ battery excitations. A physical hop across the interface changes $q$ by at most one, so the Hamiltonian connects only equal or neighboring sectors. Its Hermitian coupling between $q$ and $q+1$ is
\begin{equation}
B_q(\Phi)=\Pi_{q+1}H(\Phi)\Pi_q
+\Pi_qH(\Phi)\Pi_{q+1}.
\label{eq:SM_Bq}
\end{equation}
The modified Hamiltonian $H_q(\lambda,\Phi)$ in Eq.~\eqref{eq:gate} scales only this coupling. Starting from the empty battery, setting $\lambda=0$ prevents access to occupations above $q$ and is equivalent to evolution projected onto $0,\ldots,q$. Setting $\lambda=1$ restores all original matrix elements. The modification stays Hermitian, conserves $Q$, and leaves $H_B$ unchanged. We use it as a numerical test; we do not assume that an experiment can tune this one coupling block independently.

Table~\ref{tab:SM_gates} gives the selected full-period phases for the three coupling strengths shown in Figs.~\figpanel{fig:feedback}{a} and \figpanel{fig:feedback}{b} of the main text. Restoring the $q=1\leftrightarrow2$ coupling changes the preferred branch between $\lambda=0$ and $0.5$. Three samples do not locate the transition or determine whether the selected phase changes continuously. Restoring $q=2\leftrightarrow3$ gives a nonmonotonic change within the positive branch.

\begin{table}[htbp]
\caption{Highest-peak flux found over $[0,2\pi)$ as an occupation-sector coupling is restored. Here $N=16$, $Jt=4$; entries are $\Phi_*/\pi$. Zero coupling blocks access to the higher sector, while unit coupling recovers the original Hamiltonian.}
\label{tab:SM_gates}
\centering
\setlength{\tabcolsep}{15pt}
\begin{tabular}{ccccc}
\hline\hline
$V/J$ & boundary & $\lambda=0$ & $\lambda=0.5$ & $\lambda=1$\\
\hline
2.0 & $1\leftrightarrow2$ & 1.461834 & 0.537967 & 0.465775\\
1.5 & $2\leftrightarrow3$ & 0.598600 & 0.596525 & 0.634376\\
\hline\hline
\end{tabular}
\end{table}

For the closed $q=1\leftrightarrow2$ boundary, the old value $0.621672\pi$ was only the positive-branch maximum. The higher full-period maximum is $1.461834\pi$. The values for the $q=2\leftrightarrow3$ test remain the best phases found over the full period.

Blocking higher occupations can also change the reduced-state spectrum in sectors that remain accessible. Table~\ref{tab:SM_ranks} gives the largest state eigenvalue in each of $q=2$ and $q=3$ near the $N=20$ reference phase. These are eigenvalues of the unnormalized reduced blocks, not their total probabilities.

\begin{table}[htbp]
\caption{Largest eigenvalue of each unnormalized reduced-state block for $q=2$ and $q=3$, at $N=20$, $V=2J$, $Jt=4$, and $\Phi=0.4689\pi$. The auxiliary model adds 32 selected vectors from $q=4$ to the $m=3$ space. These values are plotted in Figs.~\figpanel{fig:feedback}{c} and \figpanel{fig:feedback}{d}.}
\label{tab:SM_ranks}
\centering
\setlength{\tabcolsep}{15pt}
\begin{tabular}{lcc}
\hline\hline
Model & largest eigenvalue in $q=2$ & largest in $q=3$\\
\hline
Exact & 0.167811295 & 0.148106571\\
CI $m=3$ & 0.167718024 & 0.189192572\\
CI $m=4$ & 0.167811336 & 0.148123146\\
32-vector $q=4$ auxiliary & 0.167921917 & 0.148566414\\
\hline\hline
\end{tabular}
\end{table}

The $m=3$ model nearly reproduces the leading $q=2$ eigenvalue, but makes the leading $q=3$ eigenvalue too large, reversing their exact order. Including $q=4$ repairs this error. Adding only 32 vectors selected from amplitudes leaking into $q=4$ also restores the order in this example. Since passive energy pairs the largest state eigenvalues with the lowest energies globally, this ordering matters even when the eigenvalues come from different occupation sectors. It identifies a projected-state error, but does not prove that this rank change alone explains the flux shift or the TDHF failure.

\suppnote{The occupation-projected model and saved prediction tests}
\label{note:sm_ci}

The projector $P_{\le m}=\sum_{q=0}^{m}\Pi_q$ acts within the fixed-$Q$ global space. Its Hamiltonian $H_m=P_{\le m}HP_{\le m}$ is Hermitian on the retained configurations, so evolution preserves normalization. Unlike TDHF, the state retains many-body correlations among these configurations. After tracing out the charger, we compute its energy and unrestricted passive energy using the same full $H_B$. No restriction on the extraction unitary is imposed by the occupation cutoff used during charging.

With the initial filled charger, $Q=N_C$. In sector $q$ there are $\binom{N_C}{Q-q}$ charger configurations and $\binom{N_B}{q}$ battery configurations. The retained dimension is therefore
\begin{equation}
D_m=\sum_{q=0}^{m}
\binom{N_C}{Q-q}\binom{N_B}{q},\quad
Q=N_C,
\label{eq:SM_Dm}
\end{equation}
Terms outside the allowed range of occupations are omitted. For $N_C=N_B=10$, the dimensions are $D_1=101$, $D_2=2126$, $D_3=16526$, and $D_4=60626$, compared with $D_{10}=184756$ for exact evolution. Table~\ref{tab:SM_hierarchy} gives their selected phases and exact regrets at $V=2J$, $Jt=4$, corresponding to Figs.~\figpanel{fig:prediction}{a} and \figpanel{fig:prediction}{b}.

\begin{table}[htbp]
\caption{Full-period flux choices and exact ergotropy regrets of the occupation-projected models at $N=20$, $V=2J$, and $Jt=4$. The independent numerical check searches all sampled local peaks. Increasing the retained dimension does not reduce the phase error or regret monotonically.}
\label{tab:SM_hierarchy}
\centering
\setlength{\tabcolsep}{15pt}
\begin{tabular}{lrrr}
\hline\hline
Model & Configurations & $\Phi_*/\pi$ & Exact regret/$J$\\
\hline
CI $m=1$ & 101 & 1.450748 & $1.043$\\
CI $m=2$ & 2126 & 0.472260 & $3.578\times10^{-4}$\\
CI $m=3$ & 16526 & 0.475521 & $1.398\times10^{-3}$\\
CI $m=4$ & 60626 & 0.469854 & $2.833\times10^{-5}$\\
Full exact & 184756 & 0.468904 & 0\\
\hline\hline
\end{tabular}

\end{table}

The reductions are smaller in the two saved-prediction systems. At $N_C=N_B=7$, $D_3=1716$ of $3432$ configurations are retained; at $N_C=N_B=8$, $D_3=3985$ of $12870$ are retained. These remove $50\%$ and about $69\%$ of the full spaces. For fixed $m$, $D_m$ grows polynomially with system size, but the cutoff required for accurate control may also grow. Moreover, unrestricted passive energy requires energies of the full $H_B$. A rank-$r$ state needs only its lowest $r$ energies, but finding them is not automatically efficient. The reduced basis alone does not establish an efficient algorithm for arbitrary many-body systems.

\subsection{Fixed-cutoff dimension and size dependence}

For balanced chains $N_C=N_B=L$, with $N=2L$ and a fixed positive integer $m$, Eq.~\eqref{eq:SM_Dm} becomes $D_m=\sum_{q=0}^m\binom{L}{q}^2$ for $L\ge m$. For each fixed $q$, $\binom{L}{q}=L^q/q!+O(L^{q-1})$. The leading term comes from $q=m$, whereas the exact dimension follows from the central binomial coefficient. Therefore,
\begin{equation}
\begin{aligned}
D_m&=\frac{L^{2m}}{(m!)^2}+O(L^{2m-1})=O(N^{2m}),\\
D_{\mathrm{ex}}&=\binom{2L}{L}\sim\frac{4^L}{\sqrt{\pi L}}.
\end{aligned}
\label{eq:SM_dimension_scaling}
\end{equation}
Thus a fixed occupation cutoff makes the propagation dimension polynomial, while the full fixed-$Q$ dimension is exponential. This is a basis-counting result for arbitrary balanced $N$, not a guarantee that the cutoff needed for accurate flux selection stays fixed or that the full unrestricted-ergotropy calculation has polynomial cost.

We separately test size dependence without changing the physical couplings or charging time. Table~\ref{tab:SM_same_parameter_sizes} uses $N_C=N_B=N/2$, $K=0.65J$, $V=2J$, $Jt=4$, $\omega_C=J$, and $\omega_B=4J$ throughout. Each projected model is evolved independently and selects a phase over $[0,2\pi)$; each regret is evaluated with exact evolution. The $N=12$--$18$ values are from new nested 64- and 128-point full-period searches with local refinement. The $N=20$ values are the independently verified hierarchy already reported in Table~\ref{tab:SM_hierarchy}. This is a systematic finite-size check at fixed parameters, not evidence for the thermodynamic limit.
\begin{table}[htbp]
\caption{Controlled size comparison with the same Hamiltonian parameters and charging time for every balanced chain. Selected phases are divided by $\pi$, and $\mathcal R_m$ denotes the exact ergotropy regret at the flux selected by cutoff $m$, in units of $J$. All maxima are numerical search results.}
\label{tab:SM_same_parameter_sizes}
\centering
\setlength{\tabcolsep}{12pt}
\begin{tabular}{rrrrrr}
\hline\hline
$N$ & $\Phi_*^{\mathrm{ex}}/\pi$ & $\Phi_*^{m=3}/\pi$ & $\mathcal R_2/J$ & $\mathcal R_3/J$ & $\mathcal R_4/J$\\
\hline
$12$ & $0.488007$ & $0.487345$ & $0.00114$ & $1.28\times 10^{-5}$ & $3.61\times 10^{-8}$\\
$14$ & $0.476112$ & $0.475196$ & $0.000281$ & $3.15\times 10^{-5}$ & $1.24\times 10^{-7}$\\
$16$ & $0.465775$ & $0.473872$ & $0.00019$ & $0.00186$ & $2.97\times 10^{-9}$\\
$18$ & $0.468867$ & $0.476611$ & $0.000219$ & $0.00188$ & $3.46\times 10^{-5}$\\
$20$ & $0.468904$ & $0.475521$ & $0.000358$ & $0.0014$ & $2.83\times 10^{-5}$\\
\hline\hline
\end{tabular}
\end{table}

\subsection{Control accuracy and state accuracy}
Let $P_{>m}=\langle\psi|(I-P_{\le m})|\psi\rangle$ be the exact probability outside the retained space at a given flux. For any normalized state $|\chi_m\rangle$ in that space, the Cauchy--Schwarz inequality bounds its squared overlap with the exact state:
\begin{equation}
|\langle\psi|\chi_m\rangle|^2
=|\langle P_{\le m}\psi|\chi_m\rangle|^2
\le\|P_{\le m}\psi\|_2^2=1-P_{>m}.
\label{eq:SM_fidelity_support}
\end{equation}
We can also bound the error in the reduced battery state itself. Write $\rho_{B,m}=\rho_B^{(m)}$ for the projected reduced state and let $Q_{>m}^{(B)}$ project onto battery occupations $q>m$. Because the projected reduced state has no weight in that range, $\Tr[Q_{>m}^{(B)}\rho_{B,m}]=0$, whereas $\Tr[Q_{>m}^{(B)}\rho_B^{\mathrm{ex}}]=P_{>m}$. The variational expression for trace distance gives, at any fixed flux,
\begin{equation}
\begin{aligned}
\frac12\|\rho_B^{\mathrm{ex}}-\rho_{B,m}\|_1
&=\max_{0\le M\le I}\big|\Tr[M(\rho_B^{\mathrm{ex}}-\rho_{B,m})]\big|\\
&\ge P_{>m}.
\end{aligned}
\label{eq:SM_battery_trace_lower}
\end{equation}
This applies to the reduced state used to calculate ergotropy, not only to the global wavefunction, and agrees with Eq.~\eqref{eq:battery_trace_lower} of the Letter. At $N=20$, $V=2J$, $Jt=4$, and $\Phi=0.4689\pi$, we find $P_{>2}\simeq0.41604$. Thus the reduced-state trace distance from any $q\le2$ model is at least $0.41604$, and the squared global overlap is at most $0.58396$. Nevertheless, the projected model selects a phase with regret only $3.578\times10^{-4}J$. Both bounds compare states at the same flux, whereas regret compares choices of flux.

Errors in stored and passive energy can partly cancel when selecting a maximum. A more accurate state therefore need not give a better phase at each cutoff. The projection preserves coherent evolution and correlations within its subspace, but is not a low-order expansion controlled by a small parameter throughout the tested range. It limits simultaneous battery occupation, not the number of hops or returns.

\subsection{Local sensitivity and nonmonotonic control regret}

The retained spaces satisfy $\mathcal H_m\subset\mathcal H_{m+1}$, but independent projected evolution does not impose a variational ordering on the resulting ergotropy curves. In particular, nesting alone does not order the selected fluxes or their exact regrets. At a fixed charging time, define the flux-dependent work error
\begin{equation}
\delta_m(\Phi)=W_B^{(m)}(\Phi)-W_B^{\mathrm{ex}}(\Phi).
\label{eq:SM_local_work_error}
\end{equation}
Let $\Phi_*^{\mathrm{ex}}$ be a smooth, isolated, nondegenerate maximum of the exact work curve, so that $\partial_\Phi W_B^{\mathrm{ex}}(\Phi_*^{\mathrm{ex}})=0$ and $\partial_\Phi^2W_B^{\mathrm{ex}}(\Phi_*^{\mathrm{ex}})<0$. Assume that the flux-dependent part of $\delta_m$ is a sufficiently small smooth perturbation near this peak and that the projected global maximum $\Phi_*^m$ remains nearby. An additive flux-independent error does not affect this assumption because it cannot shift a maximum. Expanding the stationarity condition $\partial_\Phi W_B^{(m)}(\Phi_*^m)=0$ about $\Phi_*^{\mathrm{ex}}$ gives, to leading order,
\begin{equation}
\Phi_*^m-\Phi_*^{\mathrm{ex}}
\simeq-\frac{\partial_\Phi\delta_m(\Phi_*^{\mathrm{ex}})}
{\partial_\Phi^2 W_B^{\mathrm{ex}}(\Phi_*^{\mathrm{ex}})}.
\label{eq:SM_local_flux_shift}
\end{equation}
Taylor expansion of the exact work around its maximum then yields
\begin{equation}
\begin{aligned}
W_B^{\mathrm{ex}}(\Phi_*^m)\simeq{}&
W_B^{\mathrm{ex}}(\Phi_*^{\mathrm{ex}})
+\frac{1}{2}\partial_\Phi^2W_B^{\mathrm{ex}}(\Phi_*^{\mathrm{ex}})
(\Phi_*^m-\Phi_*^{\mathrm{ex}})^2.
\end{aligned}
\label{eq:SM_local_work_expansion}
\end{equation}
Consequently, the exact ergotropy regret of the nearby projected choice obeys
\begin{equation}
\mathcal R_m\simeq
\frac{[\partial_\Phi\delta_m(\Phi_*^{\mathrm{ex}})]^2}
{2\big|\partial_\Phi^2W_B^{\mathrm{ex}}(\Phi_*^{\mathrm{ex}})\big|}.
\label{eq:SM_local_regret}
\end{equation}
The local control error is therefore governed, to leading order, by the flux derivative of the work error and the curvature of the exact peak. An approximately constant offset can make the predicted ergotropy inaccurate while leaving the optimal flux nearly unchanged. Increasing the cutoff need not reduce the error derivative monotonically, even though the retained spaces are nested. This explains why nonmonotonic control regret is mathematically possible; it does not establish a microscopic cause for the observed $m=2,3,4$ ordering.

Equations~\eqref{eq:SM_local_flux_shift} and \eqref{eq:SM_local_regret} are perturbative, local statements. They require a smooth nondegenerate peak and a nearby selected maximum; they need not apply at a nonsmooth passive-energy crossing, a degenerate maximum, or a switch to a distant competing flux branch. In particular, the $m=2$ model selects the wrong branch at five points of the interaction--time scan, whereas $m=3$ retains the correct branch throughout the tested grid (see Note~\ref{note:sm_reconstruction}). The uniform bound below addresses full-period choices without assuming that the selected maxima remain nearby.

\subsection{Uniform work error and a control-regret guarantee}

Let $W_B^{(m)}(t,\Phi)$ be the unrestricted ergotropy obtained from the independently evolved projected state, always using the same full $H_B$ as the exact ergotropy $W_B^{\mathrm{ex}}(t,\Phi)$. At a fixed charging time, define the uniform work error
\begin{equation}
\epsilon_m(t)=\sup_{\Phi\in[0,2\pi)}
\big|W_B^{(m)}(t,\Phi)-W_B^{\mathrm{ex}}(t,\Phi)\big|.
\label{eq:SM_uniform_work_error}
\end{equation}
Let $\Phi_*^{\mathrm{ex}}$ and $\Phi_*^{m}$ maximize their respective exact and projected ergotropies over the full flux period. Add and subtract the projected work at both phases:
\begin{equation}
\begin{aligned}
\mathcal R_m
={}&\big[W_B^{\mathrm{ex}}(\Phi_*^{\mathrm{ex}})-W_B^{(m)}(\Phi_*^{\mathrm{ex}})\big]\\
&+\big[W_B^{(m)}(\Phi_*^{\mathrm{ex}})-W_B^{(m)}(\Phi_*^{m})\big]\\
&+\big[W_B^{(m)}(\Phi_*^{m})-W_B^{\mathrm{ex}}(\Phi_*^{m})\big]
\le 2\epsilon_m.
\end{aligned}
\label{eq:SM_regret_bound}
\end{equation}
The middle bracket is nonpositive by optimality of $\Phi_*^m$, while the first and last brackets are each at most $\epsilon_m$. This proof holds at any finite $N$, with no smoothness or uniqueness assumption on the maxima. For a numerically selected $\widehat\Phi_m$ with projected optimization gap $\Delta_m^{\mathrm{opt}}=\max_\Phi W_B^{(m)}(\Phi)-W_B^{(m)}(\widehat\Phi_m)\ge0$, the corresponding statement is $\mathcal R(\widehat\Phi_m)\le2\epsilon_m+\Delta_m^{\mathrm{opt}}$. In the present scans, finite grids and peak refinement do not certify that $\Delta_m^{\mathrm{opt}}=0$.

A sufficient, generally conservative bound on $\epsilon_m$ follows from the projected-state error in Eq.~\eqref{eq:SM_duhamel}. Let $\Delta_B=\varepsilon_{\max}-\varepsilon_{\min}$ be the spectral width of the complete, flux-independent $H_B$, and define
\begin{equation}
\beta_m(t,\Phi)=\int_0^t
\big\|(I-P_{\le m})H(\Phi)P_{\le m}\psi_m(s,\Phi)\big\|_2\,ds.
\label{eq:SM_uniform_beta}
\end{equation}
The partial trace cannot increase trace distance, and the pure-state trace-distance identity gives
$\|\rho_B^{\mathrm{ex}}-\rho_B^{(m)}\|_1\le2\|\psi-\psi_m\|_2\le2\beta_m(t,\Phi)$.
For any two normalized battery states $\rho$ and $\sigma$, the stored-energy difference obeys
$|\Tr[(\rho-\sigma)H_B]|\le\Delta_B\|\rho-\sigma\|_1/2$.
The same bound holds for passive energy because $E_{\mathrm{pass}}(\rho)=\min_U\Tr(U\rho U^\dagger H_B)$: evaluate the difference at a minimizing unitary for either state and use the same trace inequality. Applying the triangle inequality to their difference yields
\begin{equation}
\begin{aligned}
\epsilon_m(t)&\le2\Delta_B\sup_{\Phi}\beta_m(t,\Phi),\\
\mathcal R_m&\le4\Delta_B\sup_{\Phi}\beta_m(t,\Phi).
\end{aligned}
\label{eq:SM_regret_leakage_bound}
\end{equation}
These bounds supply a sufficient condition for accurate flux selection at arbitrary finite $N$, but the required supremum has not been certified in our numerical study. Moreover, the bounds can be loose even when the actual regret is small, because accurate maximizer selection need not require uniformly accurate states or work curves. They do not establish a universal cutoff or a polynomial-time algorithm.

\subsection{Convergence of ergotropy and occupation-boundary leakage}

To examine how the independently evolved occupation cutoff approaches the
exact charging dynamics, we return to the balanced $N=16$ system with
$K=0.65J$, $V=2J$, and $Jt=4$. We evaluate the unrestricted work curves
$W_B^{(m)}(\Phi)$ at $m=1,2,3,4$ on the same 128-point flux grid
$\mathcal F_{128}=\{2\pi k/128:k=0,\ldots,127\}$. We distinguish the
sampled maxima
\begin{equation}
\begin{aligned}
\widehat\epsilon_m&=\max_{\Phi\in\mathcal F_{128}}
  |W_B^{(m)}(\Phi)-W_B^{\mathrm{ex}}(\Phi)|,\\
\widehat\beta_m&=\max_{\Phi\in\mathcal F_{128}}
  \beta_m(t,\Phi)
\end{aligned}
\label{eq:SM_sampled_cutoff_errors}
\end{equation}
from the continuous suprema in Eqs.~\eqref{eq:SM_uniform_work_error} and
\eqref{eq:SM_regret_leakage_bound}. In the leakage integral of
Eq.~\eqref{eq:SM_uniform_beta}, only the $q=m\leftrightarrow m+1$
block contributes: the Hamiltonian cannot change battery occupation by
more than one. We compute the integrand from the \emph{same} projected
trajectory used for the ergotropy at that flux. The values of
$\mathcal R_m$ are obtained from independently refined, full-period
choices and are evaluated using exact many-body evolution.

\begin{table}[htbp]
\caption{Cutoff convergence at $N_C=N_B=8$, $K=0.65J$, $V=2J$, and
$Jt=4$. The work-error and leakage maxima are over 128 sampled phases,
not certified continuous suprema. $P_{>m}$ is the exact probability of
occupations excluded by the cutoff at the exact optimum. Every regret
uses exact unrestricted ergotropy.}
\label{tab:SM_CI_convergence}
\centering
\setlength{\tabcolsep}{11pt}
\begin{tabular}{crrrrr}
\hline\hline
$m$ & $D_m$ & $\widehat\epsilon_m/J$ & $\widehat\beta_m$ & $\mathcal R_m/J$ & $P_{>m}$\\
\hline
1 & 65 & $3.518$ & $4.406$ & $0.7494$ & $0.8779$\\
2 & 849 & $0.6980$ & $2.389$ & $1.903\times10^{-4}$ & $0.4116$\\
3 & 3985 & $0.08023$ & $0.8857$ & $1.856\times10^{-3}$ & $0.06976$\\
4 & 8885 & $0.007975$ & $0.1891$ & $2.962\times10^{-9}$ & $0.003363$\\
\hline\hline
\end{tabular}
\end{table}

The sampled ergotropy error and projected boundary leakage both decrease at every cutoff, while the regret does not. Notably, $m=2$ excludes $41\%$ of the exact global-state probability at the exact optimum, yet selects a flux with regret $1.90\times10^{-4}J$. From $m=3$ to $m=4$, the work error decreases by about tenfold while regret falls by over five orders of magnitude. Thus, improving the work-curve approximation need not improve the selected control monotonically.

We also evaluated the \emph{actual reduced-state trace-norm error} at the exact optimum, obtaining $1.7924$, $0.9200$, $0.1850$, and $0.01158$ for $m=1,2,3,4$, respectively. Each respects the lower bound $2P_{>m}$ from Eq.~\eqref{eq:SM_battery_trace_lower}. In particular, the battery state becomes almost five times closer in trace norm from $m=2$ to $m=3$ while control regret increases by nearly a factor of ten. This is a direct demonstration that state and control accuracies need not have the same cutoff dependence.

For a separate check of the leakage integral, refining the Simpson time
mesh from 65 to 257 points changes the largest sampled values by less
than $10^{-7}$ in relative terms. The work curves of $m=2,3,4$ agree
with the independently archived calculations within
$2.1\times10^{-14}J$; direct exact propagation agrees with its archived
values to within $8\times10^{-15}J$ at the tested phases.
At the exact optimum, the measured global-state distances
$\|\psi-\psi_m\|_2$ are approximately $1.448$, $0.940$, $0.360$, and
$0.0722$ for $m=1,2,3,4$; each lies below its independently integrated
$\beta_m$, as required by Eq.~\eqref{eq:SM_duhamel}.

The local-regret relation in Eq.~\eqref{eq:SM_local_regret} explains why the work-curve and control errors need not have the same cutoff dependence. At the exact maximum, finite differences give $|\partial_\Phi\delta_m|/J\simeq0.0212$, $0.149$, and $0.00020$ per radian for $m=2,3,4$. Hence the slope error grows from $m=2$ to $m=3$ despite the decrease in the maximum work error, then nearly vanishes at $m=4$. The local formula predicts regrets $1.66\times10^{-3}J$ and $3.13\times10^{-9}J$ for $m=3,4$, close to their direct values. It is less accurate at $m=2$, where the local perturbation is not sufficiently small, and does not apply to the wrong-branch $m=1$ choice.

The improvement has a concrete stored-versus-passive-energy signature. Table~\ref{tab:SM_CI_slope_balance} gives their local flux slopes at the exact optimum using a five-point centered difference with step $0.001$ rad. For $m=3$ the two slopes have opposite signs, producing a substantial ergotropy slope. At $m=4$ both slopes are close and positive, and their difference is about $2\times10^{-4}J$ per radian. This near cancellation explains the very small displacement of the selected phase without requiring a uniformly accurate ergotropy curve. Changing the differentiation step between $0.0005$ and $0.002$ rad affects the smallest slopes and the $m=3$ passive slope slightly; the sign and scale of the comparison remain unchanged. The table is a numerical mechanism diagnostic, not a proof of universal cancellation.

\begin{table}[htbp]
\caption{Finite-difference flux derivatives of stored energy $E_B$, passive energy $E_{\mathrm{pass}}$, and work $W_B=E_B-E_{\mathrm{pass}}$ at the exact optimal phase, $\Phi_*^{\mathrm{ex}}\simeq0.465775\pi$, for $N_C=N_B=8$, $K=0.65J$, $V=2J$, and $Jt=4$. Values are in $J$ per radian; the derivatives of the exact work curve vanish within numerical search precision.}
\label{tab:SM_CI_slope_balance}
\centering
\setlength{\tabcolsep}{14pt}
\begin{tabular}{lrrr}
\hline\hline
Model & $\partial_\Phi E_B$ & $\partial_\Phi E_{\mathrm{pass}}$ & $\partial_\Phi W_B$\\
\hline
Exact & $+0.067495$ & $+0.067496$ & $\simeq0$\\
CI $m=2$ & $-0.138466$ & $-0.159657$ & $+0.021191$\\
CI $m=3$ & $+0.031709$ & $-0.117419$ & $+0.149128$\\
CI $m=4$ & $+0.073448$ & $+0.073244$ & $+0.000204$\\
\hline\hline
\end{tabular}
\end{table}

Finally, the analytic leakage estimate remains conservative. For the
unchanged battery Hamiltonian, $\Delta_B=46J$, and even the sampled
$m=4$ leakage inserted into the pointwise continuity estimate gives a
scale $2\Delta_B\widehat\beta_4\simeq17.4J$, much larger than the
measured $0.00798J$ work error. Moreover, a sampled maximum is not a
certified upper bound on the continuous supremum. The numerical data
therefore demonstrate fast convergence in this finite benchmark, while
the general leakage inequality supplies a sufficient, not tight,
convergence guarantee.
\begin{figure}[htbp]
\centering
\includegraphics[width=0.9\linewidth]{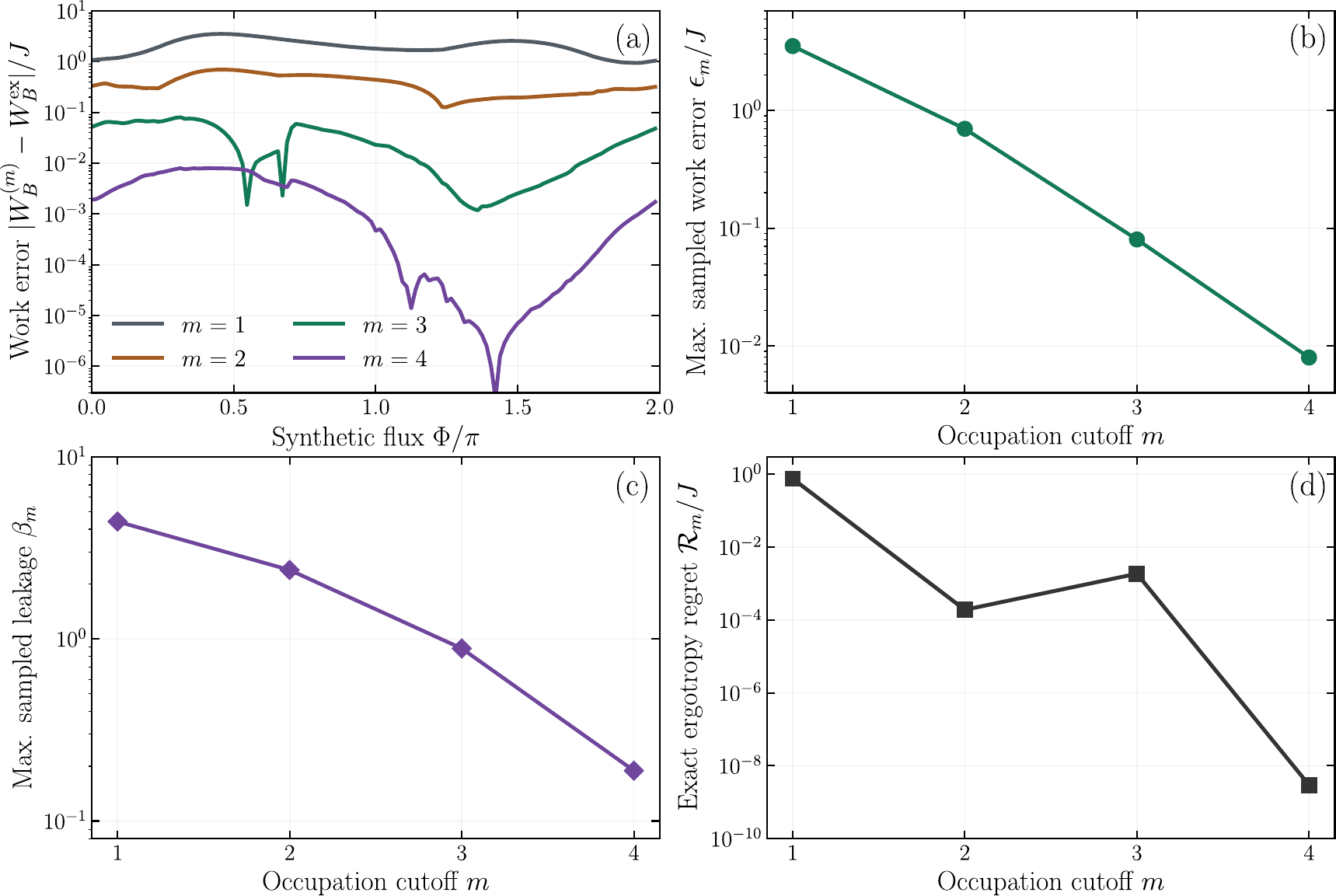}
\caption{Convergence of the occupation-projected description at
$N=16$, $K=0.65J$, $V=2J$, and $Jt=4$.
(a) Absolute ergotropy-curve error across the full flux period for
$m=1,2,3,4$; values below $10^{-7}J$ share a display floor.
(b) Largest sampled work error $\widehat\epsilon_m$.
(c) Largest sampled integrated boundary leakage $\widehat\beta_m$.
(d) Exact ergotropy regret at the independently selected flux.
The maxima in (b),(c) are over a finite 128-point grid, with all
regrets in (d) evaluated at locally refined full-period selections.}
\label{fig:SM_CI_convergence}
\end{figure}

\subsection{Predictions saved before exact evaluation}

For two parameter combinations, we saved predictions of TDHF, CI with $m=2,3$, and Gaussian reconstruction before evaluating the exact references. The prediction and exact-comparison records carry matching SHA-256 digests, which verify file identity and document the calculation sequence but do not independently authenticate creation times or constitute external preregistration. Exact references were obtained from a coarse flux grid followed by refinement of sampled peaks.

\begin{table}[htbp]
\caption{Original positive-branch choices saved before exact evaluation, shown in Figs.~\figpanel{fig:prediction}{c} and \figpanel{fig:prediction}{d}. Phases are divided by $\pi$, and regrets are evaluated with exact unrestricted ergotropy in units of $J$. These records are kept unchanged and separate from later full-period checks.}
\label{tab:SM_holdout}
\centering
\setlength{\tabcolsep}{15pt}
\begin{tabular}{lrrrr}
\hline\hline
& \multicolumn{2}{c}{$N=14$, $V=1.8J$, $Jt=4.6$}
& \multicolumn{2}{c}{$N=16$, $V=2.3J$, $Jt=4.35$}\\
Predictor & $\Phi/\pi$ & Regret/$J$ & $\Phi/\pi$ & Regret/$J$\\
\hline
TDHF & 0.781037 & 1.113467 & 0.683403 & 1.374933\\
CI $m=2$ & 0.475830 & 0.023410 & 0.403000 & 0.000720\\
CI $m=3$ & 0.500919 & 0.00001518 & 0.407205 & 0.000000433\\
Gaussian of CI $m=3$ & 0.479907 & 0.016564 & 0.387855 & 0.014594\\
Exact reference & 0.501536 & 0 & 0.407312 & 0\\
\hline\hline
\end{tabular}
\end{table}

The $m=3$ selections nearly coincide with the exact references in both cases. The later full-period checks confirm the exact, TDHF, and $m=2,3$ choices to the reported precision. They do change the $N=16$ Gaussian-of-CI choice to $1.538579\pi$, with exact regret $0.536152J$. The original $0.387855\pi$ choice and $0.014594J$ regret therefore remain positive-branch comparisons, not full-period Gaussian results. At $N=14$, the Gaussian-of-CI selection remains $0.479907\pi$. We do not replace the original saved predictions retrospectively.

\suppnote{Numerical checks, fixed-flux error bounds, and limits}
\label{note:sm_accuracy}

Exact references use the full fixed-$Q$ space and a Hermitian Hamiltonian. We check conjugate hopping amplitudes, total excitation number, state normalization, and positivity and unit trace of the reduced battery state. Passive energy always uses energies from all sectors of the unchanged $H_B$.

Reported optimal phases are numerical search results. The original two prediction tests used 17 equally spaced reference samples followed by refinement of sampled local maxima. Their small regrets are measured relative to those reference optima. The later full-period checks below are additional tests, not changes to the saved predictions.

The physical flux domain is $[0,2\pi)$. In the real occupation basis, $H(-\Phi)=H(\Phi)^*$ and the initial state is real, so $|\psi(t,-\Phi)\rangle=|\psi(-t,\Phi)\rangle^*$. Flux reversal is therefore related to time reversal, not to evolution at the same positive time. We assume no even-flux symmetry. The independently calculated $N=20$ ergotropies at $\pi/2$ and $3\pi/2$ are indeed different.

The verification solver uses nested periodic grids of 64 and 128 points for the two prediction systems and the additional stress tests, and 32 and 64 points for the $N=20$ hierarchy. It refines each sampled local maximum, including peaks across the periodic seam, and subdivides refinement brackets to resolve nearby peaks. Table~\ref{tab:SM_period} shows convergence of the exact reference values. The $N=20$, $m=1$ optimum becomes $1.45075\pi$ over the full period; the old $0.63658\pi$ value was a restricted-domain maximum. Finite grids and local searches still cannot rule out every unsampled narrow peak.

\begin{table}[htbp]
\caption{Convergence of exact-reference full-period searches. Parameters are the two prediction benchmarks of Table~\ref{tab:SM_holdout} and the $N=20$ benchmark of Table~\ref{tab:SM_hierarchy}. Each sampled local peak is refined. Agreement between grids is a numerical check, not a proof of the continuous global optimum.}
\label{tab:SM_period}
\centering
\setlength{\tabcolsep}{15pt}
\begin{tabular}{rrrr}
\hline\hline
$N$ & Grid points & $\Phi_*^{\mathrm{ex}}/\pi$ & $W_{B,*}^{\mathrm{ex}}/J$\\
\hline
14 & 64 & 0.501535973 & 7.36479924711\\
14 & 128 & 0.501535971 & 7.36479924711\\
16 & 64 & 0.407311724 & 7.49182044299\\
16 & 128 & 0.407311738 & 7.49182044299\\
20 & 32 & 0.468904136 & 7.75219643663\\
20 & 64 & 0.468904110 & 7.75219643663\\
\hline\hline
\end{tabular}

\end{table}

Additional checks compare the derivative of $\mathcal E_G$ with $F_{\mathrm{HF}}$ under a Hermitian variation, verify total-energy conservation, and confirm agreement of exact, TDHF, and exact-data Gaussian work at $K=V=0$. TDHF uses DOP853 with relative tolerance $2\times10^{-10}$ and absolute tolerance $2\times10^{-12}$. Exact observables are compared after one exponential action and four successive actions. The slope decomposition is checked with phase steps $0.002$, $0.001$, and $0.0005$ radians. These are floating-point consistency checks, not interval-arithmetic bounds.

\subsection{A bound at two fixed fluxes}

A separate estimate bounds the error of a normalized projected state without using the exact state as input. The exact and projected states start from the same initial condition. Writing the difference of their propagators in integral form gives the Duhamel bound
\begin{equation}
\|\psi(t)-\psi_m(t)\|_2
\le\int_0^t\|(I-P_{\le m})H P_{\le m}\psi_m(s)\|_2\,ds.
\label{eq:SM_duhamel}
\end{equation}
The integrand is the amplitude per unit time that the projected evolution would send outside its retained space. It depends on the projected trajectory and the full Hamiltonian, not on the unknown exact trajectory. Let $\beta_m$ bound the right-hand side. If $\bm s$ and $\bm s^{(m)}$ are the ordered Schmidt singular values of the exact and projected charger--battery states, then
\begin{equation}
\|\bm s-\bm s^{(m)}\|_2\le\beta_m.
\label{eq:SM_singular_bound}
\end{equation}
This bound does not require gaps between the Schmidt values. Combining it with the ascending energies of the full $H_B$ gives bounds on passive energy; using the same Hamiltonian also bounds stored energy. For $N=20$, $V=2J$, and $Jt=4$, the archived $q\le5$ calculation with 81 time samples gives
\begin{equation}
1.664237J\le W_B(\pi/2)-W_B(\pi)\le2.750955J.
\label{eq:SM_fixed_flux_bound}
\end{equation}
The exact difference $2.207838J$ lies inside this interval. The analytic estimates allow for values between the time samples, but the floating-point propagation values at those samples were not enclosed by interval arithmetic. The interval is therefore conditional on their numerical accuracy, not a formal computer-assisted proof. Under that assumption it establishes the sign of the difference between two fixed fluxes, not the location of a continuous optimum.

The results concern finite chains and the stated parameter ranges. The correlation-source equation is an exact hierarchy relation, not a closed kinetic theory. Increasing the occupation cutoff need not improve the selected phase monotonically. Neither a universal flux rule, a rigorously certified continuous optimum, nor a thermodynamic-limit result follows from these tests.

\suppnote{Number-conserving extraction and the scope of the work measure}
\label{note:sm_operational}

Let $Q_B=\sum_{j\in B}n_j$. Unrestricted ergotropy allows any battery unitary in a cyclic extraction process, whose initial and final battery Hamiltonian is $H_B$. A unitary that changes $Q_B$ requires excitation exchange with a control or work-storage system. The ideal work is the battery energy decrease; implementation costs are not subtracted.

If allowed unitaries instead satisfy $[U_B,Q_B]=0$, they have the block form $U_B=\bigoplus_qU_q$. Let $p_{qa}^{\downarrow}$ be the descending eigenvalues of the unnormalized block $\rho_B^{(q)}$ and $\varepsilon_{qa}^{\uparrow}$ the ascending energies of $H_B^{(q)}$. Minimizing the final energy separately in each sector gives the number-conserving ergotropy
\begin{equation}
W_B^{\mathrm{num}}=
E_B-\min_{[U_B,Q_B]=0}\Tr(U_B\rho_B U_B^\dagger H_B)
=E_B-\sum_q\sum_a p_{qa}^{\downarrow}\varepsilon_{qa}^{\uparrow}.
\label{eq:SM_number_ergotropy}
\end{equation}
Because the allowed unitaries form a smaller set, $0\le W_B^{\mathrm{num}}\le W_B$. The on-site energy $\omega_BQ_B$ cancels between the initial and minimum final energies, since $Q_B$ cannot change. For a Gaussian state, the product weights in Eq.~\eqref{eq:SM_G_weights} are grouped by $\sum_\ell n_\ell=q$ before sorting.

Table~\ref{tab:SM_number_ergotropy} reoptimizes every method for this same restricted objective over the full period. The $m=3$ choices remain accurate in both examples, but the exact preferred phase can change when the extraction constraint changes. The comparison does not assume that every number-conserving many-body unitary is easy to implement.

\begin{table}[htbp]
\caption{Full-period flux choices for number-conserving extraction in the two systems of Table~\ref{tab:SM_holdout}. Every method maximizes $W_B^{\mathrm{num}}$, not unrestricted work. The final two columns are exact regrets for this restricted objective, in units of $J$.}
\label{tab:SM_number_ergotropy}
\centering
\setlength{\tabcolsep}{15pt}
\begin{tabular}{crrrrrr}
\hline\hline
$N$ & $W_{B,*}^{\mathrm{num}}/J$ & $\Phi_*^{\mathrm{ex}}/\pi$ & $\Phi_*^{\mathrm{HF}}/\pi$ & $\Phi_*^{m=3}/\pi$ & TDHF regret & CI regret\\
\hline
14 & 3.327896 & 0.501640 & 0.797739 & 0.500554 & $0.2723$ & $7.610\times10^{-6}$\\
16 & 4.253646 & 0.356468 & 0.653258 & 0.355733 & $0.49041$ & $6.092\times10^{-6}$\\
\hline\hline
\end{tabular}

\end{table}

The main study uses unrestricted ergotropy with respect to the fixed internal $H_B$. It does not include the costs of connecting or disconnecting the charger, implementing extraction controls, or completing repeated cycles. Those costs are outside the present flux-selection comparison.

\suppnote{Simple comparison controls and twelve additional parameter tests}
\label{note:sm_validation}

A useful flux selector should be compared with simple choices as well as TDHF. Table~\ref{tab:SM_simple_controls} uses the two saved-prediction systems to compare maximizing the $m=3$ work, maximizing its stored energy, and always choosing $\Phi=\pi/2$. Every regret is evaluated with exact unrestricted ergotropy. The fixed phase is already near the $N=14$ optimum, but not the $N=16$ optimum. Maximizing projected stored energy selects a different phase from maximizing projected work in both cases.

\begin{table}[htbp]
\caption{Exact unrestricted-ergotropy regrets in the two original prediction systems. Subscripts 14 and 16 identify the chain size, with parameters from Table~\ref{tab:SM_holdout}. The work choices are the saved predictions; the fixed-phase and projected-energy controls were evaluated later.}
\label{tab:SM_simple_controls}
\centering
\setlength{\tabcolsep}{15pt}
\begin{tabular}{lrrrr}
\hline\hline
Selector & $\Phi_{14}/\pi$ & $R_{14}/J$ & $\Phi_{16}/\pi$ & $R_{16}/J$\\
\hline
TDHF & 0.781037 & $1.113$ & 0.683403 & $1.375$\\
CI $m=3$, work & 0.500919 & $1.518\times10^{-5}$ & 0.407205 & $4.331\times10^{-7}$\\
Fixed $\Phi=\pi/2$ & 0.500000 & $9.445\times10^{-5}$ & 0.500000 & $0.2478$\\
CI $m=3$, energy & 0.484036 & $0.01126$ & 0.393022 & $7.871\times10^{-3}$\\
\hline\hline
\end{tabular}

\end{table}

We then chose twelve further parameter combinations before computing their exact reference curves. They cover total sizes $12$--$16$, equal and unequal charger--battery sizes, $K/J\in\{0.35,0.65,0.95\}$, $V/J\in\{0,1,1.5,2,2.5,3\}$, and $Jt$ from 3 to 7. This is a fixed set of tests, not a random sample or every combination of the listed values. No case is removed because of its result.

For every case we first compute the $m=3$ work choice, the $m=3$ energy choice, and the TDHF choice. The complete prediction file is saved and fixed before any exact reference in the set is evaluated. Its SHA-256 digest checks file integrity but does not independently authenticate the time of creation.

Let $\widehat\Phi$ be a selected phase and $\Phi_*^{\mathrm{ref}}$ the numerical exact reference maximum. The relative exact regret is
\begin{equation}
r(\widehat\Phi)=
\frac{W_B^{\mathrm{ex}}(\Phi_*^{\mathrm{ref}})-W_B^{\mathrm{ex}}(\widehat\Phi)}
{W_B^{\mathrm{ex}}(\Phi_*^{\mathrm{ref}})}.
\label{eq:SM_relative_regret}
\end{equation}
The denominator is positive in every test. Table~\ref{tab:SM_stress} reports all twelve cases. The accompanying archive contains the scan data, local-peak lists, and saved predictions. Every method uses the same full $H_B$ and unrestricted extraction. The cutoff stays at $m=3$; it is not adjusted after seeing the exact result.

\begin{table}[htbp]
\caption{All twelve preselected parameter tests. The final four columns give relative exact ergotropy regrets in percent, $100r$. CI selects the largest $m=3$ work; energy selects the largest $m=3$ stored energy; fixed uses $\Phi=\pi/2$. Other parameters are those of Note~\ref{note:sm_hamiltonian}. No case is omitted.}
\label{tab:SM_stress}
\centering
\setlength{\tabcolsep}{9pt}
\begin{tabular}{lccccccrrrr}
\hline\hline
Case & $N_C$ & $N_B$ & $K/J$ & $V/J$ & $Jt$ & $\Phi_*^{\mathrm{ex}}/\pi$ & CI (\%) & TDHF (\%) & Energy (\%) & Fixed (\%)\\
\hline
S01 & 6 & 6 & 0.35 & 1 & 3 & 0.97050 & $1.84\times10^{-5}$ & $1.75\times10^{-3}$ & $0.612$ & $22.9$\\
S02 & 6 & 6 & 0.65 & 2 & 4 & 0.48801 & $2.10\times10^{-4}$ & $16.5$ & $0.102$ & $0.0657$\\
S03 & 6 & 6 & 0.95 & 3 & 5.5 & 0.23795 & $0.218$ & $5.81$ & $0.0156$ & $19.4$\\
S04 & 5 & 7 & 0.65 & 1.5 & 4.5 & 0.49707 & $1.74\times10^{-3}$ & $3.55$ & $0.216$ & $1.64\times10^{-3}$\\
S05 & 7 & 5 & 0.65 & 2.5 & 4.5 & 1.56655 & $4.02\times10^{-4}$ & $32.1$ & $13.8$ & $16.2$\\
S06 & 6 & 8 & 0.35 & 2 & 4 & 0.58597 & $9.05\times10^{-5}$ & $4.23$ & $0.818$ & $1.52$\\
S07 & 8 & 6 & 0.95 & 2 & 5 & 1.61621 & $3.68\times10^{-3}$ & $9.52$ & $0.0359$ & $3.09$\\
S08 & 7 & 7 & 0.95 & 1 & 5.5 & 0.54662 & $0.568$ & $9.3$ & $3.23$ & $0.774$\\
S09 & 7 & 7 & 0.35 & 3 & 3.5 & 0.16245 & $7.22\times10^{-6}$ & $20.4$ & $0.0157$ & $21.7$\\
S10 & 8 & 8 & 0.65 & 1.5 & 6 & 0.53342 & $2.43\times10^{-3}$ & $1.07$ & $6.43\times10^{-3}$ & $0.355$\\
S11 & 6 & 6 & 0.65 & 0 & 4 & 0.93739 & $2.63\times10^{-3}$ & $0.0384$ & $3.01\times10^{-4}$ & $28.6$\\
S12 & 7 & 7 & 0.65 & 2 & 7 & 0.38514 & $0.0225$ & $19.9$ & $0.211$ & $8.02$\\
\hline\hline
\end{tabular}

\end{table}

The median relative regrets are $0.0021\%$ for the $m=3$ work choice, $7.56\%$ for TDHF, $0.157\%$ for the projected-energy choice, and $5.56\%$ for the fixed phase. The largest $m=3$ regret is $0.568\%$ in S08. It beats TDHF in all twelve cases, but the energy choice is better in S03 and S11, and the fixed phase is better in S04. These exceptions are part of the comparison: the tests do not establish exact control recovery or uniform superiority over simpler choices.

\subsection{Reproducing the numerical workflow}

The archived \texttt{repro/run\_stress\_tests.py} defines all twelve cases, computes the three predicted choices, saves their records and an aggregate SHA-256 digest, and only then calculates the exact references. Each reference record contains the digest of the prediction file. The independent driver \texttt{repro/recompute\_stress\_tests.py} repeats this prediction-first sequence in a fresh directory without replacing the original records. This makes the order reproducible, although a local checksum is not external proof of chronology.

The script \texttt{verify\_archive.py} checks input hashes, prediction/reference links, and archived stress-test and robustness summaries. The script \texttt{repro/audit\_stress\_endpoints.py} independently recomputes work at saved selected phases, without rerunning every optimization. The original \texttt{generate\_all\_figures.py} reproduces four main and four original Supplemental vector-PDF figures. The companion numerical archive includes separate code and data for Fig.~\ref{fig:SM_K_robustness} and Fig.~\ref{fig:SM_CI_convergence}; the public repository currently includes the $K$-scan scripts and aggregate summary, while the new cutoff-convergence addendum is supplied separately. The saved predictions, retrospective tests, and later scans remain distinct records; the repository does not claim an independent reconstruction of every historical workflow.

\end{document}